\documentclass[12pt]{iopart}

\usepackage{hyperref}
\hypersetup{
colorlinks=true,
linkcolor={blue},
citecolor={blue},
urlcolor={blue},
bookmarksnumbered=true,
bookmarksopen=true,
breaklinks=true,
hypertexnames=false,
}

\usepackage{graphicx}
\usepackage{braket}

\usepackage{siunitx}
\DeclareSIUnit\bar{bar}

\usepackage[numbers,sort&compress]{natbib}

\newcommand{\mtext}[1]{\mathrm{#1}}

\begin{document}

\title[Electrical post-fabrication tuning of aluminum Josephson junctions at room temp.]{Electrical post-fabrication tuning of aluminum Josephson junctions at room temperature}

\author{Christian Kri\v{z}an$^{*,1}$,
Maurizio Toselli$^{*,1}$\footnote{Present address: Silicon Quantum Computing Pty Ltd., UNSW Sydney, New South Wales, Australia},
Irshad Ahmad$^1$,\\
Hadi Khaksaran$^1$,
Marcus Rommel$^1$,
Nermin Trnjanin$^1$,\\
Janka Biznárová$^1$,
Mamta Dahiya$^1$,
Emil Hogedal$^1$,\\
Halld\'{o}r Jakobsson$^1$,
Andreas Nylander$^1$,
Jonas Bylander$^1$,\\
Per Delsing$^1$,
Giovanna Tancredi$^1$}

\address{$^1$ Department of Microtechnology and Nanoscience,\\Chalmers University of Technology, SE-412 96 Gothenburg, Sweden}
\eads{\mailto{krizan@chalmers.se} and \mailto{christian.krizan@gmail.com}}
\vspace{10pt}
\begin{indented}
\item[] {\footnotesize * These authors contributed equally to this work.\\}
\item[] May 2026
\end{indented}

\begin{abstract}
Josephson junctions are key elements of superconducting quantum technology, serving as the core building blocks of superconducting qubits. We present an experimental study on room-temperature electrical tuning of aluminum junctions, showing that voltage pulses can controllably increase their resistance and adjust the Josephson energy while maintaining qubit quality factors above $1$ million. We find that the rate of resistance increase scales exponentially with pulse amplitude during manipulation. Following the manipulation, the resistance increases spontaneously through contributions proportional to both the initial resistance and the preceding manipulation. We show that this spontaneous increase halts at cryogenic temperatures, and resumes again at room temperature. Using our stepwise protocol, we achieve up to a 270\% increase in junction resistance, corresponding to a reduction of nearly \SI{2}{\giga\hertz} of the qubit transition frequency. These results establish the achievable range, relaxation behavior, and practical limits of electrical tuning, enabling post-fabrication mitigation of frequency crowding in quantum processors.
\end{abstract}
\vspace{1pc}

\section{Introduction} \label{section:introduction}

Quantum processors promise to tackle computational tasks far beyond those of classical computers, with superconducting architectures \cite{Blais2021} leading the way. Yet, their practical realization remains constrained by persistent fabrication challenges in the nanoscopic precision required to produce reliable qubits. The key element of any superconducting qubit is the Josephson junction \cite{Josephson1962, Anderson1963}, most commonly fabricated by connecting two superconducting leads by a thin dielectric oxide. The qubit transition frequencies are determined by junction properties such as its normal state resistance and its reactive environment, and can thus be engineered during the design stage. Despite advancements in the fabrication of Josephson junctions in terms of improvements in lithography, junction deposition, oxidation, evaporation, and angular deposition techniques \cite{Muthusubramanian2024, Moskalev2023, Osman2023, Kreikebaum2020, Pishchimova2023}, the state-of-the-art junction resistance spread is on the order of $2$\% \cite{Osman2024_thesis} across a \SI{50}{\milli\meter} wafer, corresponding to a frequency spread of about $\pm$\SI{50}{\mega\hertz} for transmon qubits. Frequency imprecision translates into spectral collisions between quantum operations \cite{ibmpatent2018}, that is, frequency crowding. For processors with as few as $\mathord{\sim}30$ qubits, frequency crowding can reduce the probability of collision-free operations to below single-digit percentages \cite{Osman2023, Hertzberg2021}. Scaling further to $1000$ qubits is projected to demand frequency placement with only a few \SI{}{\mega\hertz} of tolerance \cite{Hertzberg2021}. These fabrication constraints motivated the development of post-fabrication tuning techniques for Josephson junctions, where junction properties are modified to reach the desired qubit frequencies.

Post-fabrication tuning of qubit frequencies can be achieved by modifying the reactive environment of the junction, for example, through the modification of nearby capacitive elements \cite{ibmpatent2021flipchip}. An alternative approach, which is less dependent on engineering the surrounding electromagnetic reactance, is to directly tune the junction resistance. Almost universally, methods for modifying its resistance rely on heating \cite{Pavolotsky2011, Lehnert1992}. Thermal annealing below \SI{250}{\celsius} is generally believed to increase the resistance by growing the oxide layer using already-present chemisorbed oxygen \cite{Pavolotsky2011, Lehnert1992} or hydroxyl groups \cite{Pavolotsky2011}, with the contribution from hydroxyl groups likely to be smaller \cite{Lehnert1992}. At higher annealing temperatures, the barrier growth is primarily driven by oxygen diffusion \cite{Pavolotsky2011}. Through thermal annealing, resistance increases up to $400$\% and $290$\% have been reported for \SI{4}{\micro\meter\squared} {\nobreak Nb-AlO$_{\text{x}}$-Nb} junctions \cite{Lehnert1992} and {\nobreak Al-AlO$_\text{x}$-Al} junctions \cite{Koppinen2007}, respectively. However, thermally annealing a whole quantum processor does not allow for selectively tuning the resistance of individual junctions.
This limitation pushed development towards laser annealing \cite{Granata2007, vanderMeer2021, ibmpatent2018}, which can achieve a tuning range of $15$\% \cite{Hertzberg2021, Zhang2022} within tens of seconds \cite{Hertzberg2021} at a success rate of about $90$\% \cite{Zhang2022} with a $0.3$\% resistance spread. Lately, electron beam heating \cite{Groves1996} has been used to anneal junctions, demonstrating lower tuning ranges of $3$\% with approximately $1.5$\% resistance spread \cite{Balaji2024}.

A rapidly growing interest is now directed towards electrical tuning, where the resistance of an individual Josephson junction is modified by applying electrical pulses to its electrodes \cite{universityofchicago2020patent, anmr2024pct_application, Pappas2024}.
Applying pulses that alternate in polarity significantly enhances the resistance increase \cite{Pappas2024}, with demonstrated changes exceeding $130$\% \cite{alegria2025abaa}.
These recent demonstrations underscore electrical tuning as a promising candidate for precise and selective post-fabrication tuning of superconducting qubits, motivating an investigation into its mechanisms, limits, and feasibility for fine-tuning quantum processors.

A practical complication of electrical tuning is the junction's relaxation effect. At room temperature, once the tuning process is stopped, the junction resistance does not remain at its tuned value. Instead, it increases rapidly and continues to drift with a slow decay lasting several hours. For frequency targeting in real quantum processors, this effect implies that the tuning process cannot simply be stopped when the desired resistance is reached. Therefore, reliable tuning requires a quantitative understanding of this relaxation effect.

In this work, we investigate the performance and limitations of electrical tuning on Al-AlO$_\text{x}$-Al Josephson junctions.
We first characterize the natural aging of Josephson junctions, observing a gradual increase in junction resistances over time, and show that routine resistance measurements do not affect aging.
We find that the rate of resistance change scales exponentially with the applied tuning voltage.
We demonstrate that with our tuning procedure, performed entirely at room temperature, we achieve almost 100\% resistance increase in \SI{5}{\minute}.
We investigate the effects of repeated tuning cycles and demonstrate cumulative resistance increases of up to 270\%.
We characterize the spontaneous post-manipulation resistance increase that follows the electrical tuning, and show that it scales linearly with the intentionally induced resistance change.
Finally, we assess the impact of the manipulation process on transmon qubits \cite{Koch2007} subjected to frequency tuning, and find qubit quality factors preserved above one million.

\section{Samples and experimental procedures} \label{section:methods}
We investigate electrical tuning in Al-AlO$_\text{x}$-Al Josephson junctions fabricated on intrinsic Si substrates. After chemical cleaning, an Al layer is deposited and acts as the wiring layer that is defined via optical lithography and etching. The Josephson junctions are patterned via e-beam lithography and formed with double-angle Al evaporation and in-situ oxidation. We fabricated three wafers with different oxidation doses, referred to hereafter as \textit{low-dose}, \textit{medium-dose}, and \textit{high-dose}. Each wafer contains hundreds of square junctions, with electrode widths from \SI{150}{\nano\meter} to \SI{600}{\nano\meter}. The \textit{low-dose} wafer was fabricated using the PICT process \cite{Osman2021}. For completeness, it is worth mentioning that the metal films for the \textit{low-dose} and \textit{medium-dose} junctions were deposited using a Plassys MEB550S evaporator, and the metal films for the \textit{high-dose} junctions were deposited in a Plassys MEB550SL3-UHV. The detailed fabrication processes can be found in \cite{Osman2024_thesis}. 

It is important to note that our fabrication processes include multiple baking steps after Josephson junction deposition, shown in table \ref{tab:heat_exposure}. Previous work \cite{Pappas2024} has shown that heating modifies the resistance tunability of Josephson junctions. Thermal annealing of {\nobreak Nb-AlO$_{\text{x}}$-Nb} junctions has been reported for temperatures and durations comparable to those experienced by our samples \cite{Lehnert1992}. We therefore expect post-deposition heating to influence the tuning behavior observed here, although its quantitative impact is not addressed in this work.

\subsection{Active resistance manipulation} \label{section:active_resistance_manipulation}
\begin{figure}[ht!]
    \centering
    \includegraphics[width=1.00\textwidth]{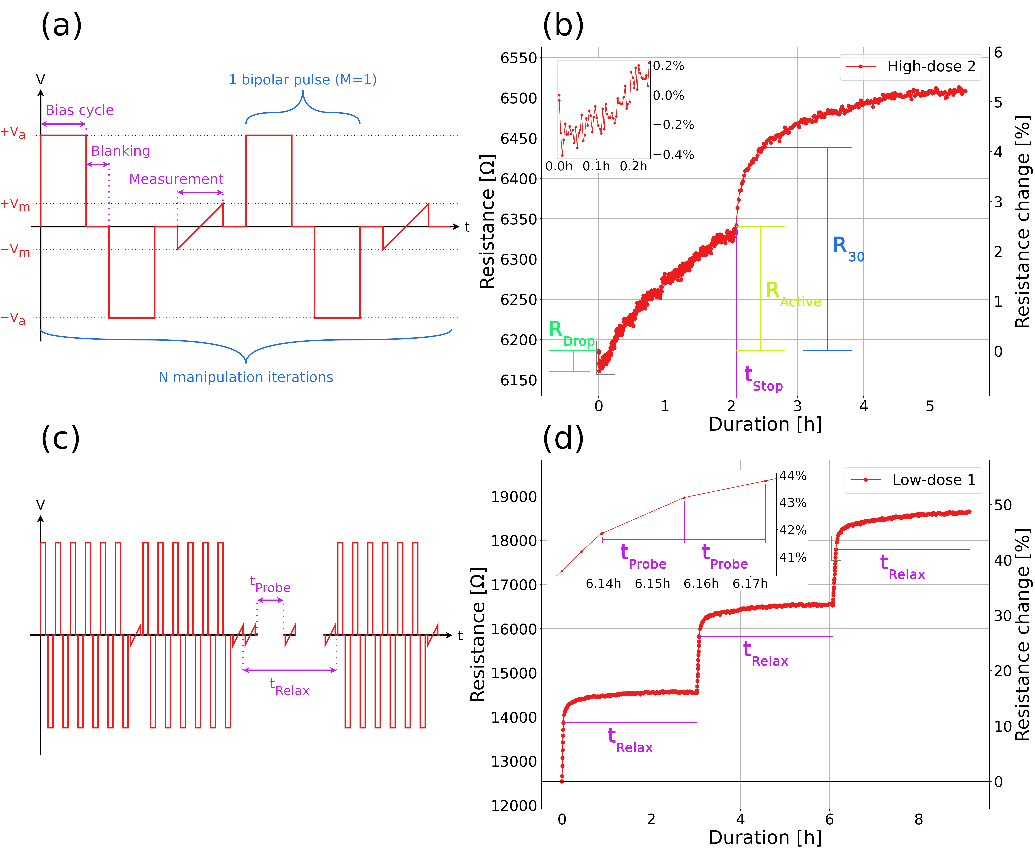}
    \caption{ \scriptsize Active resistance manipulation of a Josephson junction using two different sequences of voltage pulses. (a) Schematic illustration of the \textit{active} resistance manipulation protocol. (b) An example resistance trace of a junction manipulated up to the time $t_\mathrm{Stop}$ with the scheme in (a), showing the initial resistance drop, active manipulation, and subsequent relaxation. The inset shows a magnified view of the initial resistance drop. (c) Schematic illustration of the \textit{stepped} resistance manipulation protocol in which the \textit{active} resistance manipulation is repeated after waiting for some relaxation time $t_\mathrm{Relax}$. During relaxation, we repeatedly measure the junction's resistance with a period of $t_\mathrm{Probe}$ between measurements. (d) Measured resistance evolution during \textit{stepped} manipulation in (c). The insert in (d) highlights the beginning of the third relaxation event. The difference in initial resistance between (b) and (d) is explained by the difference in electrode width, despite the higher oxidation dose of the sample in (b). The electrode widths are \SI{354}{\nano\meter} for (b) and \SI{200}{\nano\meter} for (d). Additional sample information is provided in table \ref{tab:sample_data}.}
    \label{fig:resistance_manipulation_regions}
\end{figure}

While junctions naturally age through a passive process in which the resistance increases spontaneously over time, the resistance can also be intentionally manipulated. We refer to this process as \textit{active} resistance manipulation.
We modify the junction resistances using a bipolar voltage pulse train comprising a positive-bias square pulse, a blanking interval, a negative-bias square pulse, and a second blanking interval of equal duration. We show our standard pulse sequence in figure \ref{fig:resistance_manipulation_regions}(a), with figure \ref{fig:resistance_manipulation_regions}(b) being an example of active resistance manipulation. No substrate heating is applied during these experiments; thermal effects are discussed separately in \ref{section:thermal_process}. The protocol closely follows the alternating-bias assisted annealing technique, with minor differences in probing and measurement \cite{Pappas2024}.

As shown in figure \ref{fig:resistance_manipulation_regions}(a), each manipulation experiment comprises $N$ \textit{iterations}, each consisting of $M$ bipolar voltage pulses separated by a \SI{100}{\milli\second} blanking interval, followed by a resistance measurement. Unless otherwise noted, the pulse amplitude $\pm V_a$ is set to \SI{0.85}{\volt}, and we use $M = 6$ bipolar pulses per iteration. The duration of the square wave pulses is \SI{200}{\milli\second}, with symmetric positive and negative bias amplitudes. The resistance is extracted from an I-V measurement in which the bias voltage $V_m$ is swept linearly from \SI{-13}{\milli\volt} to +\SI{13}{\milli\volt}, while monitoring the sourced current. The junction resistance is obtained from a linear fit to the I-V data. The measurement period is intentionally set to an integer multiple of $\frac{1}{\SI{50}{\hertz}}=\SI{20}{\milli\second}$ to suppress AC mains interference. Both resistance manipulation and measurement are performed using a source-measurement unit mounted to a manual probe station in a four-wire configuration, as detailed in \ref{section:setup_for_resistance_manipulation}. This setup is used for all experiments in this work, except for the aging and dielectric breakdown characterizations reported in section \ref{section:aging_results} and \ref{section:dielectric_breakdown_results}, respectively, which are instead performed using an automated probe station as described in \ref{section:setup_for_resistance_manipulation}.

Figure \ref{fig:resistance_manipulation_regions}(b) shows the typical junction response during active manipulation. We observe three distinct regimes: (i) an initial decrease in resistance, denoted $R_\mathrm{Drop}$; (ii) the active manipulation phase, during which the resistance is intentionally increased to $R_\mathrm{Active}$; and (iii) a post-manipulation relaxation characterized by a rapid rise followed by a long-tailed saturation after the manipulation is halted at time $t_\mathrm{Stop}$. During this relaxation phase, the source-measurement unit remains connected to the junction to continuously monitor the resistance without applying manipulation pulses.

Active manipulation always carries the risk of breaking the junction, most commonly through the formation of a short-circuit across the junction.
This breaking mechanism has been linked to the glass-like behavior in {\nobreak Al-AlO$_{\text{x}}$-Al} Josephson junctions during annealing and relaxation \cite{Nesbitt2007}. This observation suggests that incorporating relaxation intervals between manipulation steps may reduce mechanical or structural strain.
For this reason, we investigated \textit{stepping} the active manipulation: we perform active resistance manipulation until a fixed increment relative to the initial resistance is reached, followed by a waiting time $t_\mathrm{Relax}$ during which the junction relaxes, after which the procedure is repeated. We refer to this protocol as \textit{stepped} active manipulation.
Figures \ref{fig:resistance_manipulation_regions}(c) and (d) illustrate the waveform used for this process, and the resulting resistance changes, respectively. The results of our stepped manipulation experiments are presented in section \ref{section:stepped_active_manipulation}.

\subsection{Effect of resistance manipulation on qubit performance}
The primary motivation of this work is to employ resistance manipulation for tuning qubit frequencies post-fabrication. We therefore investigate the impact of active manipulation on qubit decoherence and quality factor.

On the \textit{high-dose} wafer, we fabricated two nominally identical devices, $S_1$ and $S_2$, each containing eight fixed-frequency transmon qubits $Q_1 - Q_8$. $S_1$ and $S_2$ were fabricated adjacent to one another on the wafer.
We performed two cryostat cooldowns using identical device mounting positions, cabling, and instrumentation. In the first cooldown, $S_1$ served as a reference while $S_2$ underwent active resistance manipulation. Prior to the second cooldown, $S_1$ was manipulated, and $S_2$ underwent a second manipulation process. During each manipulation, the junction resistances increase, lowering the qubits' transition frequencies.

To assess qubit performance, we characterize decoherence metrics in section \ref{section:decoherence_analysis} by interleaving measurements of the energy relaxation time $T_1$, the Ramsey dephasing time $T_2^*$, and the Hahn-echoed dephasing time $T_2^e$. These measurements are repeated every \SI{15}{\minute} over approximately one day per device per cooldown, with the distribution mean reported in table \ref{tab:decoherence_data}. Qubit quality factors are extracted from the measured $T_1$ distributions.

\section{Results}
We now summarize the main results of this work. We first provide an overview of the Josephson junction samples and show that the measurement process itself does not induce resistance change, while confirming that natural aging depends on junction size.
We demonstrate room-temperature active resistance manipulation and model how the resistance manipulation depends on the amplitude of the tuning signal. We compare the total resistance change to the actively induced resistance change, and show that a stepped manipulation protocol extends the accessible resistance range. Finally, we assess the impact of resistance manipulation on the decoherence and quality factors of transmon qubits.

\subsection{Josephson junction samples}
Table \ref{tab:sample_data} summarizes parameters of the Josephson junction samples studied in this work.
$R_W$ denotes the mean initial room-temperature resistances of the unmanipulated junctions.
The electrode width $\sqrt{A}$ is defined assuming junctions of square cross-sections, and $R_W \cdot A$ is the corresponding resistance-area product. The oxidation parameters are the pressure $p_{ox}$, time $t_{ox}$, and dose $D_{ox}$, with $D_{ox}$ being calculated using the empirical formula in \cite{Zeng2015}.
\textit{Age} specifies the time elapsed between fabrication and the first measurement, where $t_0$ is defined as the time of the deposition of the second junction electrode, corresponding to the junction \textit{birth time} \cite{Nesbitt2007}.

\begin{table}[ht!]
    \centering
    \caption{ \scriptsize Summary of the five Josephson junction variants fabricated on three wafers of different oxidation doses, denoted as \textit{low-dose}, \textit{medium-dose}, and \textit{high-dose}.
    \label{tab:sample_data}
    }
    \lineup
    \begin{tabular}{llllll}
    \br
                                             & \textbf{Low-}    & \textbf{Low-}    & \textbf{Medium-} & \textbf{High-}               & \textbf{High-}             \\
    \textbf{Sample}                          & \textbf{dose 1} & \textbf{dose 2} & \textbf{dose 1} & \textbf{dose 1}             & \textbf{dose 2}           \\
    \mr
    $R_W$ [\SI{}{\ohm}]                         & $11662$           & $5513$            & $11057$             & $7840$                        & $6281$                        \\
    $\sqrt{A}$ [\SI{}{\nano\meter}]                          & $200$             & $300$             & $350$               & $318$                         & $354$                         \\
    $R_W \cdot A$ [\SI{}{\nano\ohm\meter\squared}]           & $0.47$            & $0.50$            & $1.35$              & $0.79$                        & $0.79$                        \\
    Age [days]                      & $>182$          & $>359$          & $>364$            & $>102$                      & $>102$                      \\
    $p_{ox}$ [\SI{}{\milli\bar}]                          & $2$               & $2$               & $10$                & $10$         & $10$         \\
    $t_{ox}$ [\SI{}{\minute}]                           & $20$              & $20$              & $60$                & $150^{\footnotemark[1]}$  & $150^{\footnotemark[1]}$  \\
    $D_{ox}~[\text{mbar}^{0.43} \cdot \text{min}^{0.65}]$       & $9.4$            & $9.4$            & $38.5$              & $224.0$        & $224.0$        \\
    \end{tabular}
\end{table}

\subsection{Aging} \label{section:aging_results}
We characterized junction aging by monitoring the resistance spontaneously increasing under ambient laboratory conditions, with the aim of determining whether the measurement current or cadence of measurement influences aging.

Measurements were performed at room temperature using an automatic probe station, described in \ref{section:setup_for_resistance_manipulation}. 
The junction geometries studied here differ from those stated in table \ref{tab:sample_data}: the \textit{low-dose} junctions feature electrode widths of \SI{100}{\nano\meter} and \SI{500}{\nano\meter}, and the \textit{medium-dose} junctions feature widths of \SI{200}{\nano\meter} and \SI{600}{\nano\meter}, representing small and large devices, respectively.

\footnotetext[1]{The oxidation time for the \textit{high-dose} wafer is an estimate, due to a cleanroom power failure with no machine logs being saved for the actual time.}

\begin{figure}[ht!]
    \centering
    \includegraphics[width=1.00\textwidth]{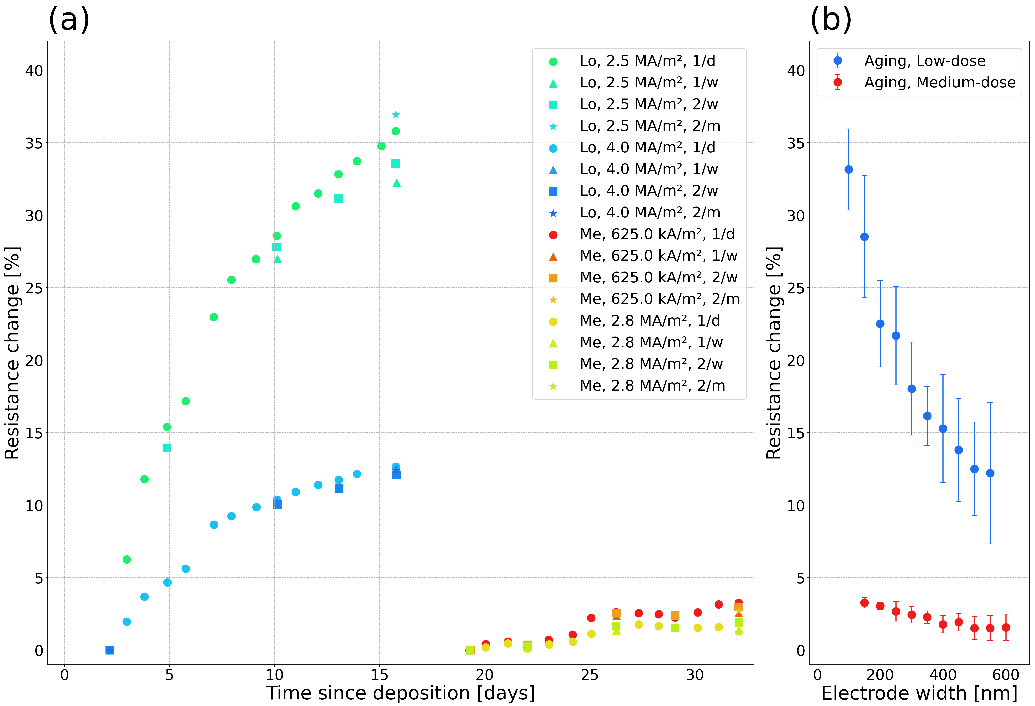}
    \caption{ \scriptsize Natural aging of \textit{low-dose} and \textit{medium-dose} junctions of different sizes, denoted as \textit{Lo} and \textit{Me}, respectively.
    (a) Resistance change $\rho(t)$ as a function of time measured at different probe currents and measurement intervals; \textit{1/d}: daily, \textit{1/w}: weekly: \textit{2/w}: biweekly, \textit{2/m}: bimonthly. The legend indicates the measurement current density $I/A$.
    (b) Resistance change $\rho(t)$ of \textit{low-dose} and \textit{medium-dose} junctions as a function of junction electrode width. The \textit{low-dose} data were measured $2$ days after fabrication, and the \textit{medium-dose} data after $19$ days. When normalized to the earliest data point, the \textit{medium-dose} junctions appear to age more slowly, likely reflecting them being $17$ days older when measured.
    }
    \label{fig:aging_junction_sizes}
\end{figure}

To characterize and quantify resistance changes we define the resistance change $\rho(t)$ as
\begin{equation} \label{eq:delta_R}
\rho(t) = \frac{R(t)}{R(0)} - 1~~,
\end{equation}
\noindent where $R(t)$ is the resistance measured at time $t$, with $R(0)$ being the initial resistance. We express $\rho(t)$ in units of $\%$ to enable comparison across junctions with different initial resistances and wafer origins. Figure \ref{fig:aging_junction_sizes}(a) shows the natural $\rho(t)$ evolution for varying measurement currents and measurement intervals; each trace corresponds to an individual junction tracked throughout the study. Similar to earlier work \cite{Bladh2005}, we observe a significant increase of $R(t)$ over a period of two weeks. We find no significant difference in aging behavior between frequently and infrequently measured samples, indicating that the resistance measurements do not measurably affect aging. Hence, we can assume that junctions can be measured safely without the junction changing its resistance beyond the extent it would have changed normally due to natural aging.
While the data suggest slower aging for \textit{medium-dose} junctions compared to \textit{low-dose} junctions, a definitive comparison is precluded by differences in the experimental conditions. In particular, measurements of the \textit{medium-dose} junctions began $19$ days after deposition, compared to $2$ days for the \textit{low-dose} junctions, and the \textit{medium-dose} junctions were additionally subjected to post-deposition heating, which is expected to mitigate subsequent aging.

Figure \ref{fig:aging_junction_sizes}(b) shows $\rho(t)$ as a function of junction size.
Each data point represents an average over measurements performed at bias currents of \SI{25}{\nano\ampere}, \SI{50}{\nano\ampere}, \SI{100}{\nano\ampere}, and \SI{1}{\micro\ampere}; for \textit{low-dose} junctions, measurements at \SI{10}{\micro\ampere} were also included. Each data point is the average of all resistance measurements on the last day in figure \ref{fig:aging_junction_sizes}(a), corresponding to ages of $16~(32)$ days for \textit{low-dose}(\textit{medium-dose}) junctions.
Independent of oxidation dose, smaller junctions exhibit faster aging. This trend is consistent with previous reports of size-dependent resistance change in Nb-Al-AlO$_\text{x}$-Nb junctions \cite{Pavolotsky2011}, although in that case the normal-state resistance was observed to decrease over time.

Overall, our observations indicate (i) the resistance spontaneously increasing by several tens of percent over a time scale of a month, (ii) faster aging for smaller junctions, (iii) no measurable influence of the measurement currents or measurement cadence on the aging process, and (iv) no conclusive difference in aging across oxidation doses within the constraints of this study.

\subsection{Active manipulation} \label{section:active_manipulation_results}
\begin{figure}[ht!]
    \centering
    \includegraphics[width=1.00\textwidth]{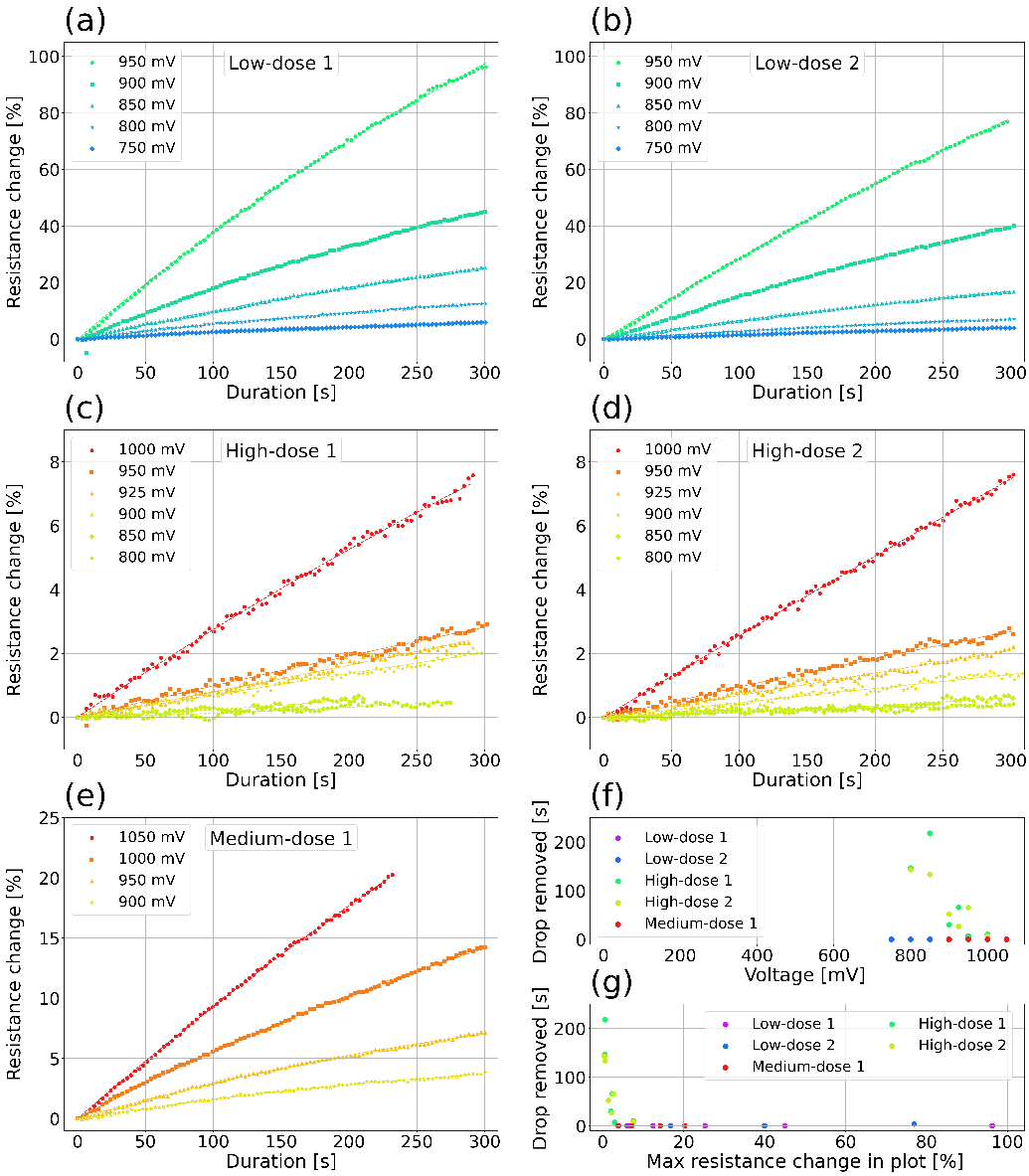}
    \caption{ \scriptsize Resistance change during active manipulation at multiple pulse amplitudes for (a) \textit{low-dose 1}, (b) \textit{low-dose 2}, (c) \textit{high-dose 1}, (d) \textit{high-dose 2}, and (e) \textit{medium-dose 1} junctions. Manipulation at voltages exceeding the highest value shown in each panel resulted in junction failure. The time-dependent resistance change is dominated by a linear term $\alpha(V)$ with a quadratic contribution $\beta(V)$, as defined in equation (\ref{eq:alpha_beta_fit_formula}). (f) and (g) show the duration of the manually excluded initial resistance drop for each device set in (a-e), as a function of the manipulation voltage (f) and the maximum resistance change (g).}
    \label{fig:active_manipulation_room_temperature}
\end{figure}

We manipulate the junction resistance by applying a train of voltage pulses, as shown in figure \ref{fig:resistance_manipulation_regions}(a).
Figures \ref{fig:active_manipulation_room_temperature}(a-e) show the actively induced $\rho(t)$ as a function of time for different pulse amplitudes $V_a$, measured across the five Josephson junctions variants listed in table \ref{tab:sample_data}. Across all devices, we find that an increasing $V_a$ yields a higher rate of resistance change over time, with a maximum increase of $96.2$\% achieved in \SI{5}{\minute}.

We fit the active resistance manipulation to a second-order polynomial equation
\begin{equation} \label{eq:alpha_beta_fit_formula}
    \rho(t) = \alpha(V) \cdot t + \beta(V) \cdot t^2~~,
\end{equation}
\noindent where $\alpha(V)$ and $\beta(V)$ quantify the linear and quadratic time dependencies, respectively. The second-order polynomial was chosen empirically based on a comparison between four different candidate models, discussed in \ref{appendix:additional_manipulation_analysis}. The fit is restricted to the active manipulation regime shown in figure \ref{fig:resistance_manipulation_regions}(b); prior to fitting, we remove the initial resistance drop, as illustrated in figures \ref{fig:active_manipulation_room_temperature}(f, g) and discussed further in \ref{appendix:resistance_drop_behavior}.

Figure \ref{fig:alpha_beta} shows the parameters obtained from fitting the data in figures \ref{fig:active_manipulation_room_temperature}(a-e) with equation (\ref{eq:alpha_beta_fit_formula}). Figure \ref{fig:alpha_beta}(a) shows the linear coefficient $\alpha(V)$ as a function of the manipulation pulse amplitude for the different samples. We find that $\alpha(V)$ exhibits an exponential dependence on voltage, which we model as
\begin{equation} \label{eq:exponential_fit_formula}
    \alpha(V) = \alpha_0 \cdot e^{V / V_0}~~,
\end{equation}
\noindent where $V_0$ is the characteristic voltage corresponding to an $e$-fold increase in $\alpha(V)$, with $\alpha_0$ being a prefactor modifying the voltage sensitivity. The extracted parameters are listed in table \ref{tab:exponential_fit_data_PERCENT}; increasing the pulse amplitude by approximately \SI{200}{\milli\volt} results in at least an order-of-magnitude increase in the resistance-change rate for the \textit{low-dose} and \textit{high-dose} junctions, whereas the \textit{medium-dose} junctions exhibit a more modest increase of about a factor of two over the measured range. Figure \ref{fig:alpha_beta}(a) also shows that samples from the same wafer cluster according to their resistance-area product $R_W \cdot A$, rather than junction area.
This observation is consistent with previous reports indicating that electrical tuning exhibits only weak dependence on junction area \cite{Pappas2024}.
Furthermore, the \textit{medium-dose} manipulation results show that $R_W \cdot A$ alone is not sufficient to predict the exponential voltage dependence. Although $\alpha(V)$ appears smaller for larger $R_W \cdot A$, the \textit{medium-dose} sample deviates from this trend, exhibiting an intermediate $\alpha(V)$ despite its comparatively large resistance-area product.
As seen in table \ref{tab:exponential_fit_data_PERCENT}, the prefactor $\alpha_0$ spans over three orders of magnitude across the junctions while 
the extracted characteristic voltage $V_0$ lies in the range \SI{55}{}$-$\SI{91}{\milli\volt}. Interestingly, these values are higher but comparable to the thermal voltage at room temperature, $V_T \equiv k_BT/e \approx \SI{25}{\milli\volt}$, consistent with slow resistance changes due to aging at room temperature. Within error bars, we also observe that $V_0$ remains constant between junction variants on a particular wafer, leaving $\alpha_0$ to capture variations in $\alpha$ between samples.

\begin{table}[ht!]
    \centering
    \caption{ \scriptsize Parameters $\alpha_0$ and $V_0$ as extracted by fitting the data in figure \ref{fig:alpha_beta}(a). The \% denotes the relative resistance change. \label{tab:exponential_fit_data_PERCENT}}
    \lineup
    \begin{tabular}{llllll}
    \br
                                             & \textbf{Low-}   & \textbf{Low-}   & \textbf{Medium-} & \textbf{High-}           & \textbf{High-}          \\
                                             & \textbf{dose 1} & \textbf{dose 2} & \textbf{dose 1}  & \textbf{dose 1}          & \textbf{dose 2}         \\
    \mr
    $\alpha_{0}$ [$10^{-9}~$\SI{}{\%\per\second}] & $824 \pm 351$        & $420 \pm 271$        & $1002 \pm 531$       & $0.79 \pm 1.19$                 & $0.37 \pm 0.37$  \\
    $V_0$ [\SI{}{\milli\volt}]               & $72.5 \pm 2.4$        & $70.4 \pm 3.4$        & $91.4 \pm 4.3$        & $57.5 \pm 5.0$                  & $55.3 \pm 3.1$                  \\
    \end{tabular}
\end{table}

Figure \ref{fig:alpha_beta}(b) shows the quadratic coefficient $\beta(V)$ extracted from equation (\ref{eq:alpha_beta_fit_formula}). The values of $\beta(V)$ are predominantly negative, consistent with a $R(t)$ increasing slower over time; occasional positive values are close to zero and are attributed to fitting uncertainty. Since the quadratic term ultimately dominates at long times, we estimate the crossover time at which $\beta(V)\,t^2$ surpasses $\alpha(V)\,t$. Across all junctions in figure \ref{fig:alpha_beta}, this time has a median of \SI{26}{\minute} and a mean of \SI{50}{\minute}, meaning that $\alpha(V)$ remains the dominant source of resistance change for the timescales observed in this study. The observation also implies the existence of a time where the resistance stops changing, assuming a hypothetical junction that survives active resistance manipulation for dozens of minutes.

Figures \ref{fig:alpha_beta}(c-e) compare $\alpha(V)$ and $\beta(V)$, and weakly suggest an anticorrelation between the two parameters, with larger $\alpha(V)$ accompanied by more negative $\beta(V)$. Physically, this trend implies that faster resistance changes are associated with more rapid deceleration of the tuning process: if a junction changes its resistance faster than before, then it would stop changing its resistance sooner. However, the limited data set precludes a statistically significant quantification of this relationship.

\begin{figure}[ht!]
    \centering
    \includegraphics[width=1.00\textwidth]{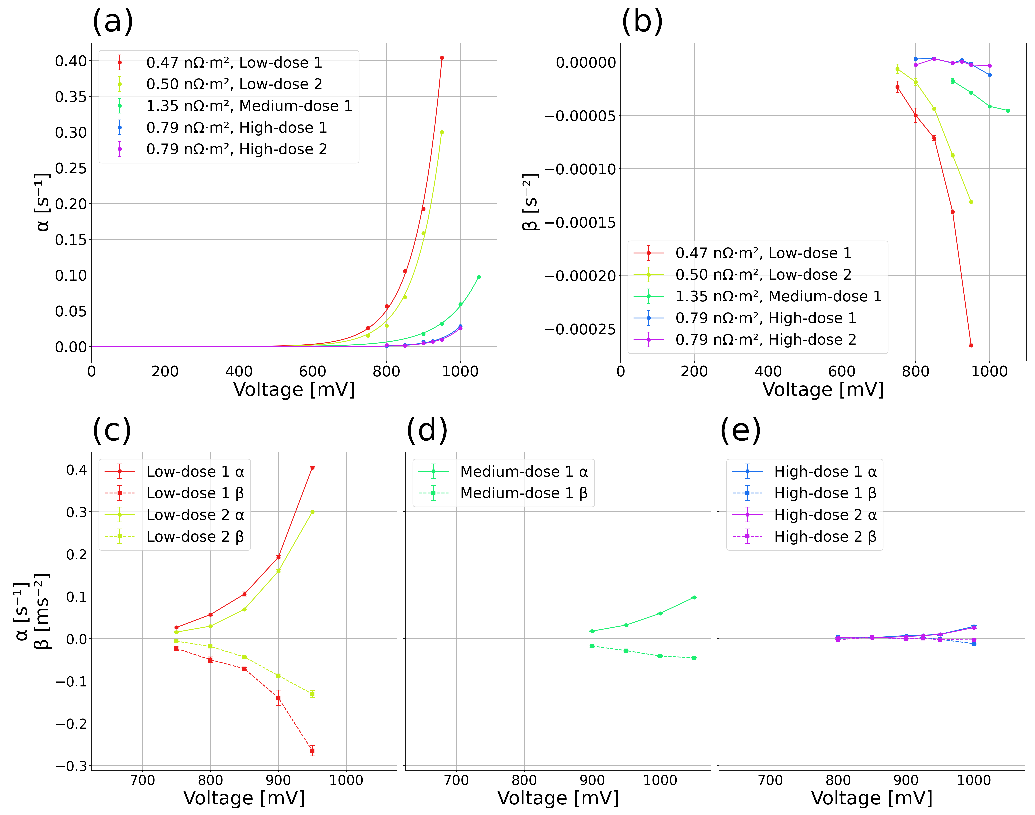}
    \caption{ \scriptsize Extracted parameters (a) $\alpha(V)$ and (b) $\beta(V)$ obtained by fitting equation (\ref{eq:alpha_beta_fit_formula}) to the resistance traces in figure \ref{fig:active_manipulation_room_temperature}. Panels (c-e) compare the voltage dependence of $\alpha(V)$ and $\beta(V)$. The linear coefficient $\alpha(V)$ exhibits an exponential dependence on voltage. The resistance-area product $R_W A~$[\SI{}{\nano\ohm\square\meter}] for the five junction variants is indicated in the legends of (a) and (b).
    }
    \label{fig:alpha_beta}
\end{figure}

\subsubsection{The initial drop in resistance}
As was shown in figure \ref{fig:resistance_manipulation_regions}(b), we observe an initial decrease in resistance at the onset of active manipulation, consistent with previous reports \cite{pappas2025patent}.
The duration and magnitude of this drop are sample-dependent. In aged \textit{low-dose} and \textit{medium-dose} junctions, the effect is typically weak and limited to one or two data points, as shown in figure \ref{fig:active_manipulation_room_temperature}(f-g). In contrast, aged \textit{high-dose} junctions exhibit a pronounced and extended resistance drop under comparable manipulation conditions.
Performing the stepped active manipulation experiment outlined in section \ref{section:active_resistance_manipulation} also amplifies the initial resistance drop between iterations of the experiment, shown in figure \ref{fig:stepped_drop_depth}.

\subsection{Relaxation dependence on active manipulation} \label{section:active_vs_total_manipulation}
Figure \ref{fig:resistance_manipulation_regions}(b) shows that the junction has increased its resistance by $R_\mathrm{Active}$ when active manipulation ceases at a time $t_\mathrm{Stop}$, corresponding to a relative change $\rho_{\mathrm{Active}}$; however, the junction's resistance continues to increase due to a relaxation mechanism up to a value $R_\mathrm{Total}$, giving a relative change $\rho_{\mathrm{Total}}$. It has been reported that, when the active manipulation is stopped, the relaxation mechanism adds a fixed fraction of the initial resistance independently of how much the resistance was increased during active manipulation \cite{pappas2025patent}. Under this observation, a plot of $\rho_{\mathrm{Total}}$ as a function of $\rho_{\mathrm{Active}}$ should yield a linear relation with unit slope and a finite offset. The offset quantifies the amount of resistance change after manipulation, independent of the active resistance change introduced during manipulation.

To test this behavior across different junction variants, we performed active manipulation to varying values of $\rho_\mathrm{Active}$, measured the corresponding $\rho_\mathrm{Total}$, and made a linear fit to the data. As the relaxation process follows a logarithmic time dependence and therefore lacks a well-defined asymptote here, we define the resistance measured 30 minutes after the manipulation is halted as \mbox{$R_{30} = R(t_{\mathrm{Stop}} + \SI{30}{\minute})$}, and consider the total resistance change $\rho_{\mathrm{Total}}(t=\SI{30}{\minute}) = \frac{R_{30}}{R(0)}-1$.

\begin{figure}[ht!]
    \centering
    \includegraphics[width=1.00\textwidth]{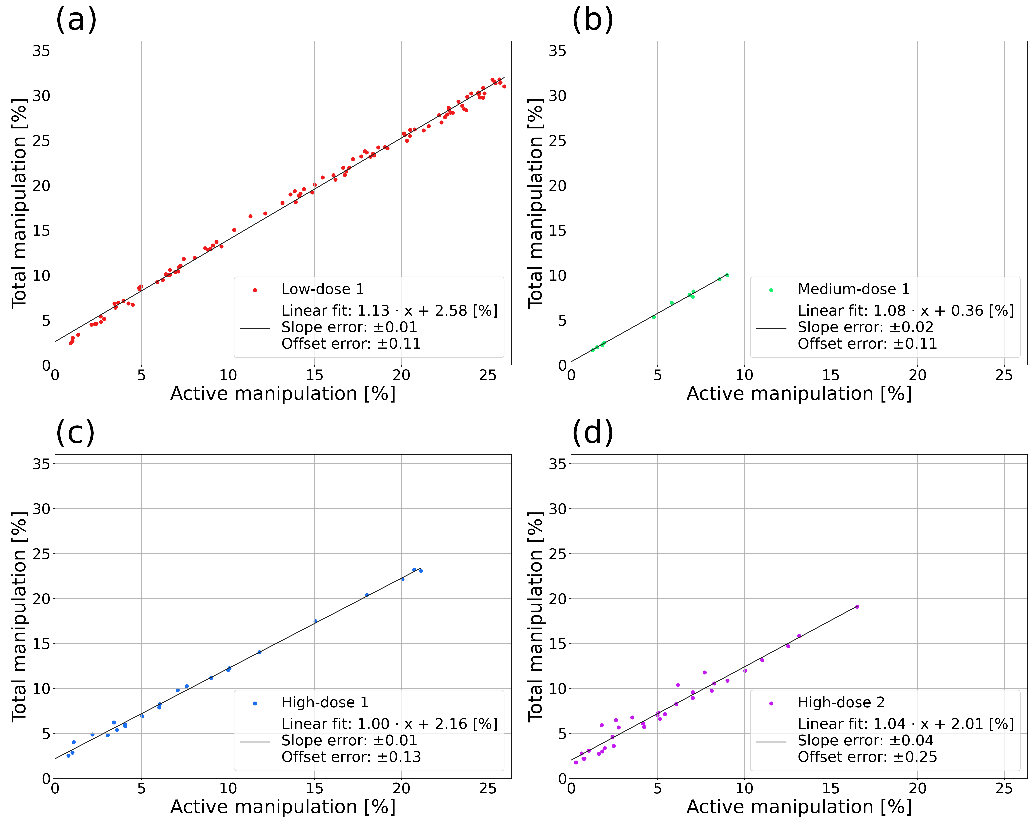}
    \caption{ \scriptsize Linear fit of the resistance change 30 minutes after manipulation, as a function of the \textit{active} resistance change for (a) \textit{low-dose 1}, (b) \textit{medium-dose 1}, (c) \textit{high-dose 1}, and (d) \textit{high-dose 2} junctions.
    Each datapoint corresponds to one independent manipulation and relaxation. A slope of unity indicates that the relaxation contributes a fixed resistance increase offset independent of the actively induced resistance change.}
    \label{fig:active_vs_total_manipulation}
\end{figure}

Figure \ref{fig:active_vs_total_manipulation} shows the resistance change $\rho_\mathrm{Total}(t)$ at $t=\SI{30}{\minute}$ as a function of the actively induced change $\rho_\mathrm{Active}$ for the (a) \textit{low-dose 1}, (b) \textit{medium-dose 1}, (c) \textit{high-dose 1} and (d) \textit{high-dose 2} junctions. The data is fitted with $\rho_\mathrm{Total}(t=\SI{30}{\minute})= k\,\rho_\mathrm{Active} + m$. The \textit{high-dose} samples reproduce the previously reported behavior, with fitted slopes $k$ of $1.00 \pm 0.01$ and $1.04 \pm 0.04$, for \textit{high-dose 1} and \textit{high-dose 2}, respectively. In contrast, \textit{low-dose 1} and \textit{medium-dose 1} exhibit larger $k$ of $1.13 \pm 0.01$ and $1.08 \pm 0.02$, respectively, indicating manipulation-dependent relaxation. We attribute these deviations in $k$ to physical effects rather than systematic error, given the goodness of fit, low scatter, and the identical measurement protocol yielding unit slopes for the high-dose devices. Moreover, the fitted offsets $m$ differ across all samples, suggesting that relaxation is governed by intrinsic material properties.

\begin{figure}[ht!]
    \centering
    \includegraphics[width=1.0\textwidth]{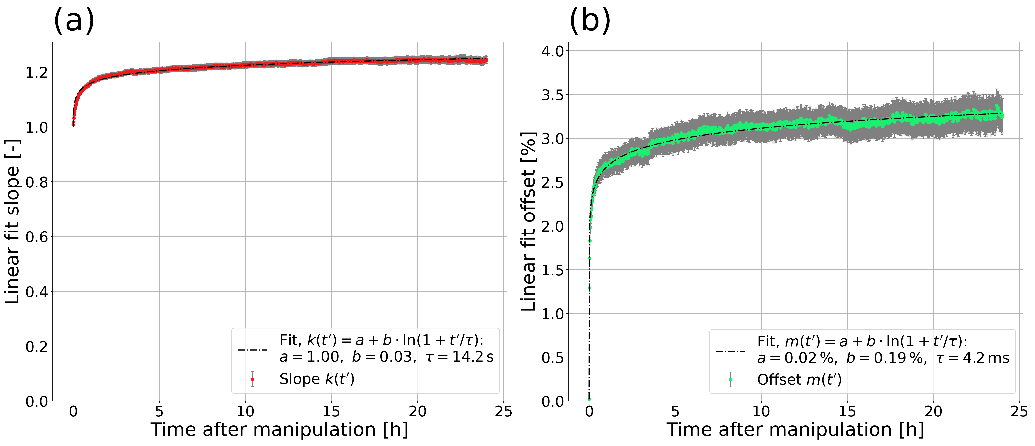}
    \caption{ \scriptsize Time dependence of the slope $k$ (a) and offset $m$ (b) extracted from linear fits at different times $\tau$ after manipulation of $\rho_\mathrm{Total}(\tau)= k\,\rho_\mathrm{Active} + m$. An example fit where $\tau=\SI{30}{\minute}$ is shown in figure \ref{fig:active_vs_total_manipulation}(a).}
    \label{fig:parameter_analysis_from_active_vs_total_manipulation}
\end{figure}

As shown in figure \ref{fig:resistance_manipulation_regions}(b), the resistance change due to relaxation is not instantaneous. While figure \ref{fig:active_vs_total_manipulation} establishes a linear relationship between $\rho_\mathrm{Active}$ and $\rho_{\mathrm{Total}}(t=\SI{30}{\minute})$, predicting the relaxation of a given junction variant requires determining whether this relationship changes over time. We therefore examined how the parameters $k$ for the slope and $m$ for the offset extracted from the linear fit between $\rho_\mathrm{Total}$ and $\rho_\mathrm{Active}$ evolve with the time elapsed after manipulation. We measure the resistance at different times, $R(t_{\mathrm{Stop}}+\tau)$, with $\tau$ ranging from \SI{0}{\hour} to \SI{24}{\hour}, and evaluate $\rho_\mathrm{Total}(\tau)=\frac{R(t_{\mathrm{Stop}}+\tau)}{R(0)}-1$. At each time step, we perform a linear fit and extracted the values of the slope $k$ and offset $m$ as a function of $\tau$. This analysis was performed on the \textit{low-dose 1} junctions, for which the largest data set was available. The results are reported in figure \ref{fig:parameter_analysis_from_active_vs_total_manipulation} and show that neither the slope nor the offset stabilizes within \SI{24}{\hour}. To be able to model the relaxation over time, we fit the time dependence of $k$ and $m$ empirically to a logarithmic growth function,
\begin{equation} \label{eq:log_fit}
    y(t) = a + b \cdot \ln\left(1 + {t}/{\tau}\right) ~~.
\end{equation}
\noindent The logarithmic model was selected over a power-law fit, following a comparison reported in \ref{appendix:log_vs_power_in_relaxation}. The extracted fit parameters are shown in the legends.

Figure \ref{fig:parameter_analysis_from_active_vs_total_manipulation}(a) shows how parameter $k$ changes as a function of the time elapsed after manipulation is halted. We find that, in addition to the fixed relaxation offset observed independently of the magnitude of active manipulation, the relaxation process contributes with approximately one quarter of $\rho_\mathrm{Active}$ to $\rho_\mathrm{Total}$.
This result demonstrates that the relaxation is not always independent of the active manipulation. The amount of $\rho(t)$ gained during relaxation, due to the amount of active resistance manipulation made, is comparable in magnitude to the frequency tolerances required for tuning quantum processors.
Figure \ref{fig:parameter_analysis_from_active_vs_total_manipulation}(b) shows that the $\rho(t)$ acquired from relaxation, regardless of the active manipulation, is $3.29 \pm 0.19\%$. The uncertainty of $\pm0.19\%$ is particularly relevant for qubit frequency targeting: as estimated in \ref{section:calculate_imprecision}, this uncertainty sets a lower bound of \SI{11.3}{\mega\hertz} on the achievable qubit frequency precision, about five times better than the regular fabrication precision.
These results have direct practical implications for post-fabrication tuning: both $k$ and $m$ of the relaxation process must be characterized to achieve accurate electrical tuning, requiring multi-day monitoring of statistically significant ensembles of sacrificial junctions.

\subsection{Stepped active manipulation} \label{section:stepped_active_manipulation}
\begin{figure}[ht!]
    \centering
    \includegraphics[width=1.00\textwidth]{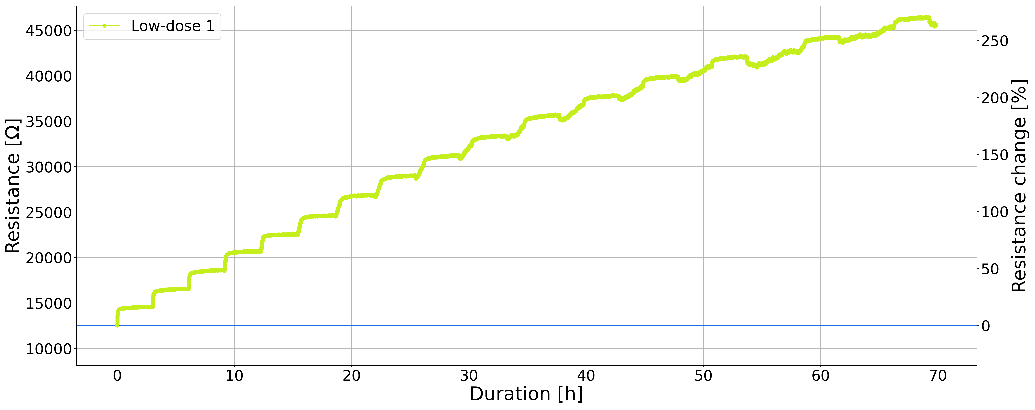}
    \caption{ \scriptsize Stepped active manipulation of a \SI{200}{\nano\meter} \textit{low-dose} junction. Each step consists of an active resistance manipulation sequence that increases the junction resistance by $+0.10\cdot R(0)$, followed by a \SI{3}{\hour} relaxation period. The steps are repeated until junction failure, achieving a resistance change of $\rho = +270\%$. Using equation (\ref{eq:full_normal_state_resistance_to_plasma_frequency}), this resistance change corresponds to a transmon frequency shift of $\Delta f = -\SI{1964}{\mega\hertz}$.}
    \label{fig:stepped_active_manipulation}
\end{figure}

We now demonstrate that stepped active manipulation can increase $\rho(t)$ beyond the 96\% achieved with regular active manipulation. The pulse sequence for the stepped scheme is shown in figure \ref{fig:resistance_manipulation_regions}(c). Figure \ref{fig:stepped_active_manipulation} presents the resistance evolution of a \textit{low-dose 1} junction, aged 348 days at the time of measurement, as it is subjected to successive manipulation steps.
Using this protocol, we increase the junction resistance by $269.5$\%, which, to our knowledge, is the largest resistance change reported for an individually targeted Josephson junction, corresponding to a transmon qubit shifting its frequency by $\Delta f = -\SI{1964}{\mega\hertz}$. Upon further stepping, the junction ultimately fails during the initial resistance drop of the final manipulation step.
We observe that the rate of resistance change decreases with each successive step. At the same time, the magnitude of the initial resistance drop increases between steps, as shown in figure \ref{fig:stepped_drop_depth}, to which we are unable to offer an explanation at this time.

\begin{figure}[ht!]
    \centering
    \includegraphics[width=1.00\textwidth]{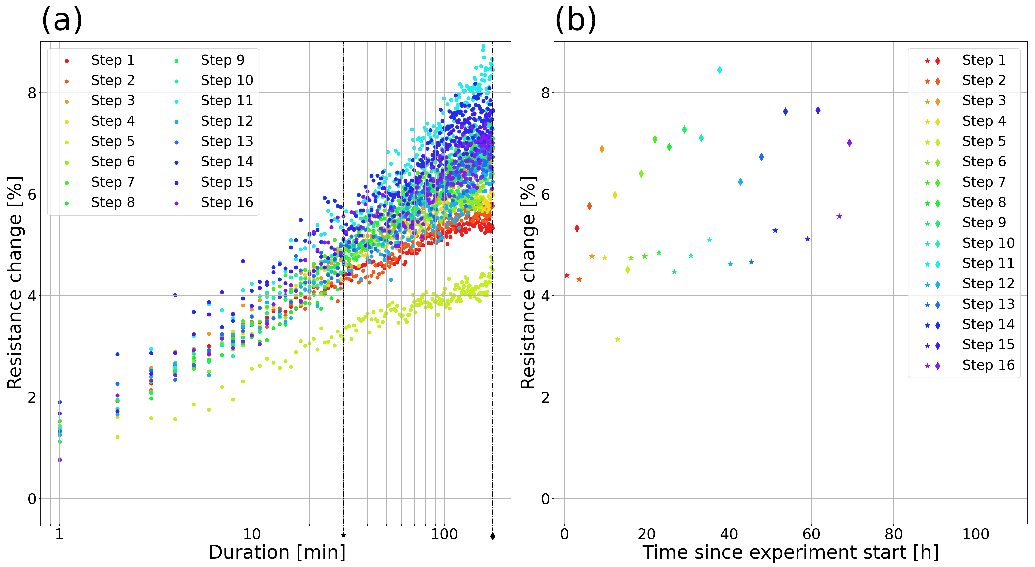}
    \caption{ \scriptsize Analysis of relaxation during the stepped active manipulation in figure \ref{fig:stepped_active_manipulation}. (a) $\text{Lin-log}_{10}$ plot of the relative resistance change $\rho$ during relaxation after each manipulation step in figure 7, confirming a logarithmic growth with time. Subsequent relaxation traces mainly follow the same time dependence as the first relaxation event, indicating repeated activation of the same relaxation process rather than a new one for each manipulation. (b) Relative resistance change $\rho$ evaluated at the \SI{30}{\minute} and \SI{180}{\minute} markers indicated in (a) as dashed vertical lines and shown as stars and diamonds, respectively. The increasing slope across steps indicates a growing relaxation rate with successive manipulations, suggesting that ongoing relaxation from previous steps accumulates.}
    \label{fig:characteristics_stepped_active_manipulation}
\end{figure}

Figure \ref{fig:characteristics_stepped_active_manipulation}(a) shows the relaxation response normalized to the initial resistance at the beginning of the experiment \mbox{$R(t=0)$}, where each trace corresponds to the relaxation phase of a single manipulation step in figure \ref{fig:stepped_active_manipulation}.
All steps exhibit similar relaxation dynamics, following an approximately logarithmic time dependence.
The main difference between steps is that the relaxation traces are progressively shifting towards higher resistance values with each successive manipulation step.
Figure \ref{fig:characteristics_stepped_active_manipulation}(b) shows the junction's resistance increase during each relaxation phase, measured at the \SI{30}{\minute} and \SI{180}{\minute} markers in figure \ref{fig:characteristics_stepped_active_manipulation}(a) as a function of time since the start of the stepped manipulation. The data reveal that the speed of the resistance increase grows with successive steps, suggesting that relaxation from earlier steps persists and contributes to subsequent relaxation phases.

\subsection{Decoherence and qubit quality factors} \label{section:decoherence_analysis}
We study the effect of resistance manipulation on qubit coherence and quality factors using two devices, $S_1$ and $S_2$, each containing eight fixed-frequency transmon qubits $Q_{1..8}$. The chips were fabricated on the \textit{high-dose} wafer, with qubits $Q_{1..4}$ having \textit{high-dose 1} junctions, and qubits $Q_{5..8}$ having \textit{high-dose 2} junctions. During the first cooldown, $S_1$ was measured as-fabricated, while $S_2$ underwent resistance manipulation. In the second cooldown, both devices were manipulated, producing a ``single-tuned'' $S_1$ and a ``double-tuned'' $S_2$. The qubits' junctions were manipulated using the manipulation procedure described in section \ref{section:active_resistance_manipulation}.

The qubit quality factors are calculated as $Q = 2 \pi f_{01} \cdot T_1$ where $f_{01}$ is the qubit frequency, and $T_1$ its energy relaxation time measured during the two cooldowns. The measured $f_{01}$, $T_1$ and calculated $Q$ values are reported in tables \ref{tab:device_parameter_table} and \ref{tab:decoherence_data}. Here, we acquire $T_1$, $T_2^*$ and $T_2^e$ using decoherence time experiments \cite{Herrmann2024_thesis}, performing interleaved $T_1-T_2^*-T_2^e$ measurements and reporting the mean values together with their uncertainties. We describe our fitting procedures in \cite{Krizan2025}.

\begin{figure}[ht!]
    \centering
    \includegraphics[width=0.50\textwidth]{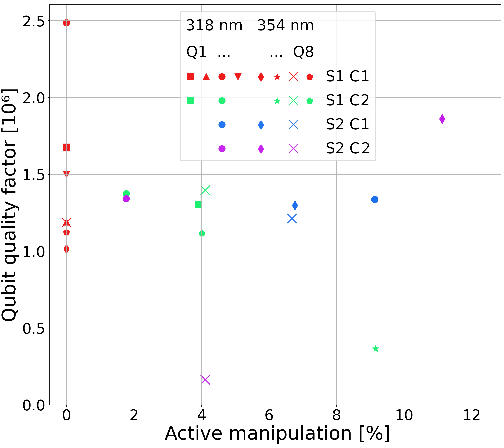}
    \caption{ \scriptsize Qubit quality factor as a function of the active resistance manipulation. Each symbol shape denotes the same qubit across measurements. With a few exceptions, the qubit quality factors lie between $1\cdot10^6$ and $1.5\cdot10^6$ for one round of manipulation, comparable to those obtained with our standard fabrication process.
    }
    \label{fig:q_factors_of_quantum_devices}
\end{figure}

Figure \ref{fig:q_factors_of_quantum_devices}(a) shows qubit quality factors as a function of $\rho_{\mathrm{Active}}$ reached prior to each cooldown, denoted \textit{C1} and \textit{C2}, respectively.
Red symbols indicate the unmanipulated reference sample $S_1$ while green symbols show $S_1$ after a single round of manipulation. Blue symbols correspond to the nominally identical sample $S_2$, manipulated once before \textit{C1}, and purple symbols indicate $S_2$ after a second manipulation round prior to \textit{C2}.
We find that a single round of resistance manipulation preserves qubit performance, with $Q$ values between $1$ and $1.5$ million, comparable to our previously reported devices \cite{Osman2023}. Data following two manipulation rounds are inconclusive, with $Q_5$ increasing in quality factor after major additional tuning, and $Q_7$ decreasing in quality factor after minor additional tuning.

While resistance manipulation was successfully performed on most qubits, a fraction broke during the procedure. Electrostatic discharge (ESD) is unlikely the cause, as manipulation was initiated and sustained on all qubits for some time. Notably, failures occurred predominantly early in the process, with approximately half of the failures occurring within the first \SI{6}{\minute} as shown in \ref{section:dielectric_breakdown_results}. We attribute the elevated failure rate primarily to fabrication issues in the \textit{high-dose} wafer. This wafer was produced using a newly commissioned evaporator for which the fabrication process had not yet been optimized; subsequent analysis revealed a less uniform deposition rate compared to other wafers studied here. Similarly, we note that $R_W \cdot A$ of the \textit{high-dose} junctions being \SI{0.79}{\nano\ohm\meter\squared}, is lower than that of the \textit{medium-dose 1} junctions, which measured \SI{0.87}{\nano\ohm\meter\squared} $19$ days after deposition. This difference can thus not be attributed to aging, and suggests that fabrication-dependent factors have contributed to either the lower $R_W \cdot A$ of the \textit{high-dose} wafer; or, the higher $R_W \cdot A$ of the \textit{medium-dose} wafer, for instance from its additional thermal exposure during fabrication. Additional dielectric breakdown data for junctions from this wafer are presented in \ref{section:dielectric_breakdown_results}. Further work is needed to distinguish whether failures stem from the manipulation protocol or wafer-specific defects.

\section{Discussion and conclusion}
In conclusion, we demonstrate a feasible method for manipulating the resistance of Josephson junctions.
By studying multiple junction variants differing in size and oxidation dose, we confirm that natural aging is junction-size dependent, with smaller junctions aging more, and that 
resistance measurements performed at intervals ranging from daily to bimonthly do not measurably affect aging.
We confirm earlier reports that the junction resistance can be actively increased at room temperature using electrical voltage pulses \cite{Wang2024}. We find that the induced increase in resistance follows a second-order time dependence dominated by its linear rate by two orders of magnitude. Crucially, we find that the linear rate depends exponentially on pulse amplitude, increasing by a factor of $e$ for every additional \SI{55}{}$-$\SI{91}{\milli\volt} depending on the sample.
Our data show that the resistance-area product alone does not predict the voltage dependence of the rate of resistance change, with one dataset disrupting an otherwise clear negative relationship.

Contrary to previous observations \cite{pappas2025patent}, the resistance gained from relaxation is not always a fixed fraction of the initial junction resistance. We find that this post-relaxation resistance change can increase with the amount of active manipulation made.
During relaxation, the amount of resistance change gained from active manipulation, and gained passively from the relaxation process, both scale linearly with the initial resistance, with the latter showing a partial dependence on the former.
We show that relaxation has a logarithmic time dependence regardless of the active manipulation, indicating that a manipulated sample at room-temperature requires monitoring over several hours to reliably extract relaxation parameters.
The resulting uncertainty limits the achievable frequency precision of electrical tuning, which we estimate to be at least \SI{11.3}{\mega\hertz}, and can possibly be reduced with greater control of the relaxation mechanism.
We introduce a stepped active manipulation scheme that extends the accessible resistance range, achieving a total increase of up to $270\%$ for an aged \SI{12.5}{\kilo\ohm} junction at room temperature, the largest targeted resistance manipulation reported to date. We find that repeated steps amplify the initial resistance drop, an effect to which we currently offer no conclusive explanation. Moreover, repeated resistance manipulation steps leave cumulative effects in subsequent relaxation phases: the resistance change measured at a fixed time after each relaxation onset, shows yet a gradual increase as the stepped manipulation experiment progresses. Finally, we show that performing resistance manipulation on qubits does not degrade their performances, preserving quality factors above $1$ million that are comparable to our standard fabrication outcomes. We do not observe any systematic dependence of qubit quality factors on the manipulation process.

As superconducting quantum processors continue to scale, techniques enabling rapid, selective post-fabrication correction become increasingly critical. Electrical tuning provides a promising route, and this work clarifies both its capabilities and its practical limitations.

\subsection{Open questions and future experiments}
Multiple open questions remain that warrant further investigation.
Future work should clarify how resistance manipulation affects junction parameters most relevant for qubit frequency targeting, namely the superconducting gap $\Delta_g(T)$, the charging energy $E_C$, and the room-temperature to cryogenic resistance ratio $R/R_N$, discussed in \ref{section:relating_resistance_to_qubit_frequency}.

While we find the rate of junction resistance increase exponentially dependent on voltage amplitude, systematic studies of other waveform parameters, such as pulse duration, asymmetry, bandwidth-induced overshoot, and shape, are warranted. Similarly, the dependence of the relaxation process on the manipulation voltage merits investigation. Complementary studies of dielectric properties before and after tuning, including breakdown voltage measurements, would provide insight into possible dielectric modifications induced by manipulation.

To assess contemporary speculations on whether the resistance manipulation is related to Cabrera-Mott oxidation, one could carefully increase the resistance until it plateaus, and subsequently apply a higher pulse amplitude to test whether resistance manipulation resumes. Theory suggests \cite{Cabrera1949} that the resistance change would stagnate once an oxide film grows to a thickness where the electric field across it becomes too weak to sustain further growth; a revived resistance increase would suggest an oxide growth driven by an electric-field intensity.
Similarly, the physical origin of the initial resistance drop remains unclear; the stepped active manipulation introduced here offers a promising route to isolating this effect as the drop increases with step count. Finally, extrinsic factors such as junction age and area should be disentangled by comparing relaxation slopes and offsets across same-wafer devices differing only in these parameters.

\section*{Data availability statement}
The data that support the findings of this study are openly available at the following URL/DOI: \url{https://doi.org/10.5281/zenodo.17817324} \cite{Zenodo_data_and_code}.

\ack
For additional insights assisting this work, we thank Daryoush Shiri for his expertise in superconductivity and Anuj Aggarwal for discussions on quantum device design.
We thank William Oliver at MIT for valuable insights and discussions regarding the possible physical mechanism behind electrical tuning and its similarities to natural aging.
We thank Jürgen Lisenfeld at KIT for helpful discussions on electrical tuning and for suggesting the experiment where a Josephson junction is rapidly submerged in liquid nitrogen during relaxation, which we present in the appendix.

We acknowledge financial support from the Knut and Alice Wallenberg Foundation (KAW) through the Wallenberg Center for Quantum Technology (WACQT), and in part from the EU Flagship on Quantum Technology under project {\nolinebreak HORIZON-CL4-2022-QUANTUM-01-SGA} grant 101113946 OpenSuperQPlus100. The devices were fabricated at the Myfab Chalmers Nanofabrication Laboratory.

\section*{Conflict of interest}
A.N. and M.R. are board members and shareholders of Arkeon Technologies AB. The other authors declare no competing interests.

\appendix
\setcounter{section}{0}

\section{Experimental setup} \label{section:setup_for_resistance_manipulation}
The room-temperature resistance manipulation and measurements were performed using a Keysight B2902B source-measurement unit, connected to a Keyfactor SimplePS manual probe station equipped with MPI MP40 probes and {\nobreak PT-W-S-2MIC} tungsten needles. All measurements were conducted in an ESD-protected area, with an Ion Systems 6430 ionizer fan continuously directed at the sample being manipulated. Aging and dielectric breakdown measurements were carried out in an MPI {\nobreak TS2000-D} automated probe station, using a Keithley 2612B System SourceMeter and a Keithley {\nobreak 3706A-S} System Switch for sample multiplexing.

\begin{figure}[ht!]
    \centering
    \includegraphics[width=0.7903\textwidth]{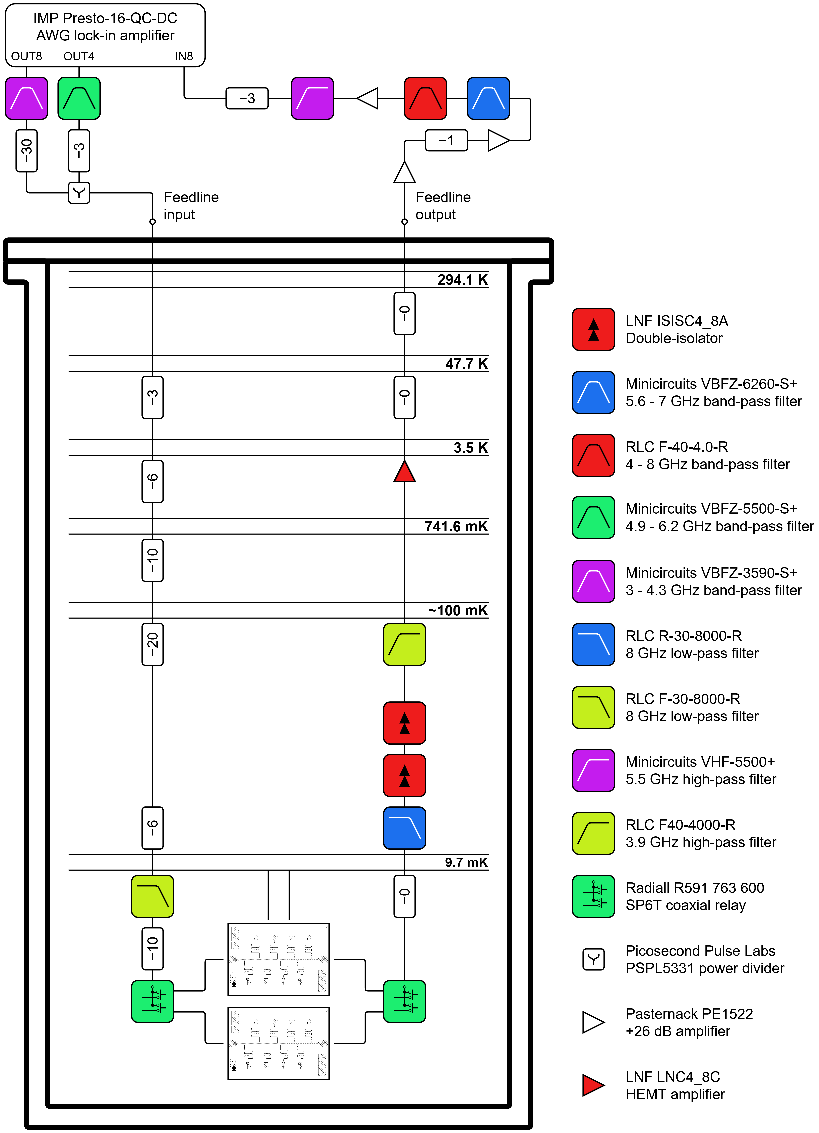}
    \caption{ \scriptsize Full wiring diagram for the experimental setup. The numbers inside the attenuators indicate the attenuation in \SI{}{\decibel}. The room-temperature attenuators before the feedline input optimize the operating range of the generated signals.}
    \label{fig:cryogenic_experimental_setup}
\end{figure}

\begin{figure}[ht!]
    \centering
    \includegraphics[width=0.50\textwidth]{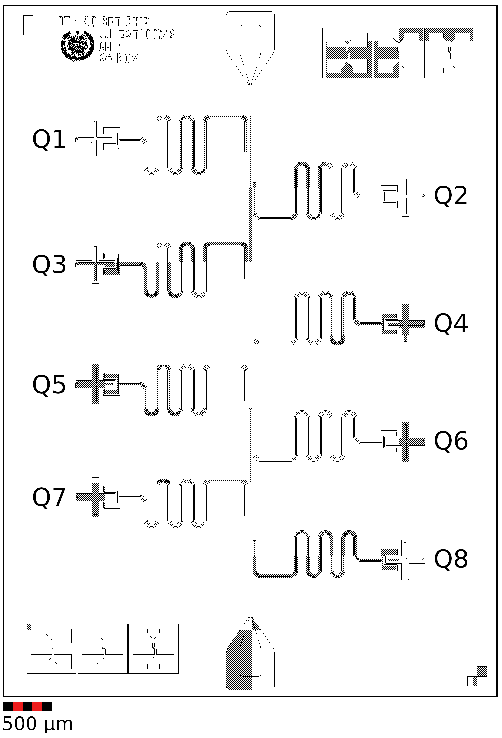}
    \caption{ \scriptsize Quantum device chip layout. Qubits $Q_1 - Q_4$ correspond to the \textit{high-dose 1} junction variant in table \ref{tab:sample_data}, while $Q_5 - Q_8$ correspond to \textit{high-dose 2} junctions. The two samples $S_1$ and $S_2$ were fabricated adjacent to one another on the same wafer.}
    \label{fig:quantum_chip}
\end{figure}

The microwave setup used for qubit characterization is illustrated in figure \ref{fig:cryogenic_experimental_setup}. The samples were thermalized at the \SI{10}{\milli\kelvin} stage of a Bluefors LD250 dilution refrigerator with highly-attenuated RF lines. Samples $S_1$ and $S_2$ share the same input/output readout line for all experiments, using coaxial microwave switches. Waveform generation and readout are handled using an IMP Presto-16-QC-DC AWG lock-in amplifier \cite{Tholen2022, Tholen2023}.
The samples, illustrated in figure \ref{fig:quantum_chip}, consist of $8$ fixed-frequency transmon qubits whose readout resonators share a common feedline. The device parameters are given in table \ref{tab:device_parameter_table}.

\section{Device parameters} \label{section:device_parameters}
Table \ref{tab:device_parameter_table} lists the measured and extracted parameters for samples $S_1$ and $S_2$ in the two cooldowns \textit{C1} and \textit{C2}. Table \ref{tab:decoherence_data} lists the acquired decoherence factors.

\begin{table}[ht!]
    \centering
    \caption{ \scriptsize Characteristic parameters of samples $S_1$ and $S_2$, which underwent one and two cycles of resistance manipulations, respectively. \textit{C1} and \textit{C2} denote the two cryostat cooldowns.
    $f_{\mathrm{res},\ket{0}}$ is the resonator frequency when the corresponding qubit is in $\ket{0}$, $\kappa$ is the energy decay rate of the resonator, $\eta$ is the transmon anharmonicity, $f_{01}$ is the qubit's $\ket{0} \rightarrow \ket{1}$ transition frequency, and $2\chi$ is the dispersive shift. \label{tab:device_parameter_table}}
    \lineup
    \begin{tabular}{lllllllllll}
    \br
    \boldmath{$S_1$} & \boldmath{$f_{\mathrm{res},\ket{0}}$} & \boldmath{$\kappa / 2\pi$} & \boldmath{$\eta / 2\pi$} &    & \boldmath{$f_{01}$} &    & \boldmath{$2\chi$} &    & \boldmath{$Q$} &    \\
                     & [\SI{}{\giga\hertz}]     & [\SI{}{\kilo\hertz}]       & [\SI{}{\mega\hertz}]     &    & [\SI{}{\giga\hertz}]    &    & [\SI{}{\kilo\hertz}]        &    & [-]                          &    \\
                     &                                       &                            & C1                       & C2 & C1                  & C2 & C1                & C2 & C1                           & C2 \\
    \mr
    $Q_1$ & $5.9532$ & $818$  &          & $-214.4$ & $4.8514$ & $4.5518$ & $-3108$ & $-1030$ & $1\,676\,534$ & $1\,302\,120$ \\
    $Q_2$ & $6.0659$ & $707$  & $-210.0$ &          & $4.9853$ &          & $-1764$ &        & $1\,190\,290$ &               \\
    $Q_3$ & $6.1935$ & $730$  &          & $-222.2$ & $4.1680$ & $4.0851$ & $-1024$ & $-411$  & $2\,487\,886$ & $1\,375\,136$ \\
    $Q_4$ & $6.3155$ & $692$  & $-210.3$ &          & $4.9750$ &          & $-1435$ &        & $1\,500\,422$ &               \\
    $Q_5$ & $6.4513$ & $767$  & $-202.3$ &          & $5.5673$ &          & $-2887$ &        & $1\,014\,434$ &               \\
    $Q_6$ & $6.5924$ & $812$ & $-204.8$ & $-197.6$ & $5.4177$ & $4.9811$ & $-1671$ & $-899$  & $1\,123\,330$ & $365\,889$    \\
    $Q_7$ & $6.7446$ & $924$ & $-201.7$ & $-207.2$ & $5.3942$ & $5.1232$ & $-1434$ & $-965$  & $1\,186\,250$ & $1\,397\,712$ \\
    $Q_8$ & $6.8918$ & $898$ & $-203.2$ & $-205.3$ & $5.5847$ & $5.2863$ & $-1627$ & $-1159$ & $1\,122\,872$ & $1\,116\,143$ \\
    \br
    \boldmath{$S_2$} &                                       &                            &                        &  &                   &  &                 &  &                            &  \\
    \mr
    $Q_1$ & $5.9509$ & $646$  & $-212.7$ &          & $4.7759$ &          & $-1495$ &        & $840\,215$    &               \\
    $Q_2$ & $6.0668$ & $647$  &          &          &          &          &        &        &               &               \\
    $Q_3$ & $6.1908$ & $754$  & $-216.8$ & $-222.0$ & $4.4353$ & $4.2354$ & $-778$  & $-620$  & $1\,337\,651$ & $1\,341\,931$ \\
    $Q_4$ & $6.3127$ & $735$  &          &          &          &          &        &        &               &               \\
    $Q_5$ & $6.4554$ & $777$  & $-207.4$ &          & $5.1634$ & $4.8590$ & $-1495$ & $-2054$ & $1\,297\,704$ & $1\,860\,994$ \\
    $Q_6$ & $6.5811$ & $735$  &          &          &          &          &        &        &               &               \\
    $Q_7$ & $6.7419$ & $815$  & $-208.9$ &          & $4.9492$ & $4.8039$ & $-857$  & $-574$  & $1\,212\,784$ & $163\,470$    \\
    $Q_8$ & $6.8838$ & $1011$ &          &          &          &          &        &        &               &               \\
    \end{tabular}
\end{table}

\begin{table}[ht!]
    \centering
    \caption{ \scriptsize Decoherence times for samples $S_1$ and $S_2$ across two cooldowns, \textit{C1} and \textit{C2}. Entries labeled \textit{n/a} indicate data with uncertainties too large to report reliably, while blank entries correspond to qubits that failed. $S_2$ was resistance-manipulated prior to \textit{C1}, whereas both $S_1$ and $S_2$ underwent manipulation prior to \textit{C2}.
    }
    \label{tab:decoherence_data}
    \lineup
    \begin{tabular}{llllllll}
    \br
    &    & \textbf{$T_{1,~\mtext{C1}}$}  & \textbf{$T_{1,~\mtext{C2}}$}   & \textbf{$T_{2,~\mtext{C1}}^*$}  & \textbf{$T_{2,~\mtext{C2}}^*$}    & \textbf{$T_{2,~\mtext{C1}}^e$}  & \textbf{$T_{2,~\mtext{C2}}^e$} \\
    &                    & [\SI{}{\micro\second}] & [\SI{}{\micro\second}]      & [\SI{}{\micro\second}] & [\SI{}{\micro\second}]      & [\SI{}{\micro\second}] & [\SI{}{\micro\second}] \\
    \mr
    \boldmath{$S_1$} & $Q_1$ & $55\pm6$ & $46\pm7$ & $25\pm6$ & $23\pm5$ & n/a & $29\pm3$ \\
                     & $Q_2$ & $38\pm7$ &  & $47\pm3$ &  & $73\pm4$ &  \\
                     & $Q_3$ & $95\pm17$ & $54\pm12$ & $34\pm16$ & $17\pm4$ & $65\pm7$ & $62\pm8$ \\
                     & $Q_4$ & $48\pm10$ &  & $49\pm13$ &  & $72\pm5$ &  \\
                     & $Q_5$ & $29\pm4$ &  & $38\pm5$ &  & $57\pm4$ &  \\
                     & $Q_6$ & $33\pm2$ & $12\pm1$ & $19\pm7$ & $4\pm1$ & $56\pm5$ & $21\pm1$ \\
                     & $Q_7$ & $35\pm5$ & $43\pm8$ & $9\pm1$ & $40\pm6$ & $57\pm7$ & $52\pm6$ \\
                     & $Q_8$ & $32\pm3$ & $34\pm7$ & $37\pm4$ & $37\pm7$ & $53\pm6$ & $44\pm6$ \\
    \mr
    \boldmath{$S_2$} & $Q_1$ & $28\pm3$ &  & $21\pm4$ &  & $37\pm3$ &  \\
                     & $Q_2$ &  &   &  &   &  &   \\
                     & $Q_3$ & $48\pm7$ & $50\pm9$ & $36\pm16$ & $21\pm2$ & $66\pm7$ & $57\pm7$ \\
                     & $Q_4$ &  &   &  &   &  &   \\
                     & $Q_5$ & $40\pm3$ & $61\pm11$ & $33\pm5$ & $14\pm10$ & $61\pm4$ & n/a \\
                     & $Q_6$ &  &   &  &   &  &   \\
                     & $Q_7$ & $39\pm5$ & $5\pm0.1$ & $39\pm8$ & $1\pm0.1$ & $67\pm9$ & n/a \\
                     & $Q_8$ &  &   &  &   &  &   \\
    \end{tabular}
\end{table}

\section{Heat exposure post-deposition}
Table \ref{tab:heat_exposure} summarizes the post-deposition thermal processing applied to the fabricated samples. The \textit{low-dose} junctions were fabricated using the PICT process \cite{Osman2021}, in which the patch-related processing steps were omitted. The \textit{medium-dose} wafer additionally included airbridge fabrication, introducing heating in the additional fabrication steps.

\begin{table}[ht!]
    \centering
    \caption{\scriptsize Summary of post-deposition heat exposure during fabrication.
    }
    \label{tab:heat_exposure}
    \lineup
    \begin{tabular}{llll}
    \br
    \textbf{Heating step} & \textbf{Temperature} & \textbf{Duration} & \textbf{Applicable samples} \\
                          & [\SI{}{\celsius}]    & [\SI{}{\minute}]  &                             \\
    \mr
    Lift-off after             &     &     &             \\
    junction deposition.        & 85  & 150 & All \\
    Resist baking for          &     &     &             \\
    patch lithography.          & 160 & 15  & Medium-dose, High-dose \\
    Patch lift-off.             & 85  & 180 & Medium-dose, High-dose \\
    Dicing-protective          &     &     &             \\
    resist baking.              & 100 & 1   & All \\
    Final resist removal.       & 85  & 15  & All \\
    Resist baking for          &     &     &             \\
    airbridge lithography.      & 115 & 1.5 & Medium-dose \\
    Resist reflow              &     &     &             \\
    for airbridges.             & 160 & 4   & Medium-dose \\
    Additional resist baking   &     &     &             \\
    for lithography and dicing. & 100 & 1.7 & Medium-dose \\
    Additional resist removal.  & 90  & 20  & Medium-dose \\
    \br
    \end{tabular}
\end{table}

\section{Relating resistance to qubit frequency} \label{section:relating_resistance_to_qubit_frequency}
The normal state resistance of a Josephson junction relates to the energy gap between the transmon qubit's $\ket{0}$ and $\ket{1}$ states through the Josephson energy \cite{Koch2007}, and ultimately through the Ambegaokar-Baratoff relation \cite{AmbegaokarBaratoff1963, errata_AmbegaokarBaratoff1963}.
For a transmon qubit with charging energy $E_C = e^2/(2\cdot C)$ where $4 E_C = \frac{(2e)^2}{2C}$ corresponds to the energy of a single Cooper pair stored in the transmon's capacitor \cite{Koch2007, Rahamim2019}, and Josephson energy $E_J$, the transition frequency $f_{01}$ can be approximated \cite{Didier2018} to
\begin{equation} \label{eq:plasma_frequency_of_transmon}
    h ~ f_{01} = \sqrt{8~E_J~ E_C} - E_C ~ (1 + \xi ~1/4 + \xi^2 ~21/128) ,
\end{equation}
\noindent where $\xi = \sqrt{2 ~ E_C / E_J}$, and $h$ is the Planck constant.
The Josephson energy $E_J$ is related to the critical current $I_C$ via
\begin{equation} \label{eq:josephson_energy_E_J}
    E_J = \frac{\Phi_0}{2 \pi} ~ I_C = \frac{\hbar}{2 e} ~ I_C ~~,
\end{equation}
\noindent where $\Phi_0$ is the magnetic flux quantum and $e$ the electron charge. The critical current $I_C$ relates to the normal state resistance $R_N$ of the junction through the Ambegaokar-Baratoff relation
\begin{equation} \label{eq:ambegaokar_baratoff}
    I_C = \frac{\pi \Delta_g(T)}{2e ~ R_N} \tanh{ \left( \frac{\Delta_g (T)}{2 k_\text{B} T} \right)} ~ ~ ,
\end{equation}
\noindent where $\Delta_g(T)$ is the superconducting energy gap, $k_\text{B}$ is the Boltzmann constant, and $T$ is the superconductor temperature. For a non-manipulated \textit{medium-dose} junction, we measured
$\Delta_g(T \approx 10~\text{mK}) = 174.3$ \textmu eV,
yielding $\tanh{ \left( \frac{\Delta_g (T)}{2 k_\text{B} T} \right)} \approx 1$ in equation (\ref{eq:ambegaokar_baratoff}). We note that our measured superconducting energy gap is close to the textbook number for bulk aluminum $\Delta_g(T) \approx 178.2$ \textmu eV \cite{Mangin2017}, resulting in {\nobreak $\frac{2\Delta_g(0.01)}{\alpha_{\text{BCS}} \cdot k_\text{B}} = T_c \approx~$\SI{1.16}{\kelvin}}, where $\alpha_{\text{BCS}} = 3.5$ is the ideal gap ratio \cite{Bardeen1957}. For reference, the nominal film thicknesses of this junction are \SI{39}{\nano\meter} for the bottom electrode, and \SI{78}{\nano\meter} for the top electrode.

Substituting equation (\ref{eq:ambegaokar_baratoff}) into equation (\ref{eq:josephson_energy_E_J}), and subsequently into equation (\ref{eq:plasma_frequency_of_transmon}), completes a relation between the normal state resistance $R_N$ and the qubit frequency $f_{01}$,
\begin{equation} \label{eq:full_normal_state_resistance_to_plasma_frequency}
    h  f_{01} = \sqrt{8 E_C \frac{\hbar \pi \Delta_g(T)}{4e^2  R_N} } - E_C\left(1+\frac{1}{4}\sqrt{\frac{2E_C}{\frac{\hbar\pi \Delta_g(T)}{4e^2  R_N} }}+\frac{21}{128}\sqrt{\frac{2E_C}{\frac{\hbar \pi \Delta_g(T)}{4e^2  R_N} }}\right).
\end{equation}

$R_N$ relates to the room temperature Josephson junction's resistance $R$. We use a fixed conversion rate $R_{N} = R \cdot 1.1385$ based on previous measurements taken on a \textit{low-dose} sample \cite{Osman2024_thesis}. Hence, we can use equation (\ref{eq:full_normal_state_resistance_to_plasma_frequency}) to estimate the change in qubit frequency given an induced change in $R$.
We measure the junction's resistance from its I-V characteristics using a voltage sweep from \SI{-13}{\milli\volt} to +\SI{13}{\milli\volt}, a region where the junction is resistor-like \cite{Toselli2024_thesis}, fitting the data to a purely linear model $V = R\cdot I$ giving us a high coefficient of determination $r^2 = 1.000(1)$.

\subsection{Tuning frequency precision from the relaxation offset error} \label{section:calculate_imprecision}
In figure \ref{fig:parameter_analysis_from_active_vs_total_manipulation}(b), the uncertainty in the relaxation fit offset relating $\rho_{\mathrm{Total}}$ to $\rho_{\mathrm{Active}}$, is approximately $\pm0.19\%$. This uncertainty directly translates into an imprecision in the qubit frequency, which we estimate here.

We conservatively assume that the relaxation-offset uncertainty extracted from the \textit{low-dose 1} junctions represents an equal or worse case for both \textit{high-dose} variants used as qubit devices, for which we have experimental data.
We further assume an energy gap $\Delta_g(T \approx \SI{10}{\milli\kelvin}) = 174.3$ \textmu eV, measured on a \textit{medium-dose} junction, and use the room-temperature to normal-state resistance conversion $R_N = R \cdot 1.1385$, as described in the previous section.
We base our estimate for frequency precision on the unmanipulated reference device's \textit{high-dose 2} junctions, using qubits $Q_5 - Q_8$ in table \ref{tab:device_parameter_table}, which have a mean transition frequency $f_{01} = \SI{5.4910}{\giga\hertz}$ and a mean anharmonicity $\eta/2\pi = \SI{-203.0}{\mega\hertz}$. Using \cite{Didier2018} with the same $\xi$ from equation (\ref{eq:plasma_frequency_of_transmon}), $E_C$ is given by
\begin{equation} \label{eq:anharmonicity_to_E_C}
    \frac{\eta}{2\pi} = -E_C -\frac{9}{16} (E_C~ \xi) -\frac{81}{128} (E_C ~ \xi^2) -\frac{3645}{4096} (E_C ~ \xi^3) -\frac{46899}{32768} (E_C ~ \xi^4) ~ .
\end{equation}
Enforcing $\eta/2\pi = \SI{-203.0}{\mega\hertz}$ in equation (\ref{eq:anharmonicity_to_E_C}) while simultaneously solving (\ref{eq:full_normal_state_resistance_to_plasma_frequency}) to $f_{01} = \SI{5.4910}{\giga\hertz}$, creates a bound solution for $R$. Numerically, we find $R$ to be \SI{5521.4}{\ohm}, resulting in $E_J = \SI{21.63}{\giga\hertz}$ from equations (\ref{eq:josephson_energy_E_J}) and (\ref{eq:ambegaokar_baratoff}), and $E_C = \SI{186.7}{\mega\hertz}$ from equation (\ref{eq:anharmonicity_to_E_C}).

We repeat this procedure for $R \pm0.19 \%$, which yields a frequency error of $5496.6 - 5485.3 = \SI{11.3}{\mega\hertz}$. We therefore assume that the uncertainty associated with the relaxation offset alone imposes a best-case bound of approximately $\SI{11.3}{\mega\hertz}$ on the attainable frequency precision of electrical tuning, under the assumption considered here.

\section{Additional manipulation analysis with secondary trends} \label{appendix:additional_manipulation_analysis}
This section details additional analysis and observations used to refine the interpretation of resistance changes in our devices.
We first address the removal of the initial resistance drop observed in figure \ref{fig:active_manipulation_room_temperature}, which is necessary for fitting the time-dependent resistance using equation (\ref{eq:alpha_beta_fit_formula}).
We then justify the use of a second order polynomial in equation (\ref{eq:alpha_beta_fit_formula}) to model active resistance manipulation.
Finally, we supplement figure \ref{fig:alpha_beta} and table \ref{tab:exponential_fit_data_PERCENT} with equivalent results in units of \SI{}{\ohm\per\second} and \SI{}{\ohm\per\second\squared}.

\subsection{Resistance drop} \label{appendix:resistance_drop_behavior}
The resistance drop is removed from the dataset prior to fitting equation (\ref{eq:alpha_beta_fit_formula}). For completeness, the removed resistance drop for each fit is shown as a function of manipulation voltage in figure \ref{fig:active_manipulation_room_temperature}(f), and as a function of the maximum reported resistance change in figure \ref{fig:active_manipulation_room_temperature}(g).
We observe no clear dependence of the initial resistance drop on either resistance change or voltage range. The only systematic trend is that \textit{high-dose} devices consistently exhibit a significantly larger drop than the \textit{low-dose} and \textit{medium-dose} devices.

\begin{figure}[ht!]
    \centering
    \includegraphics[width=0.50\textwidth]{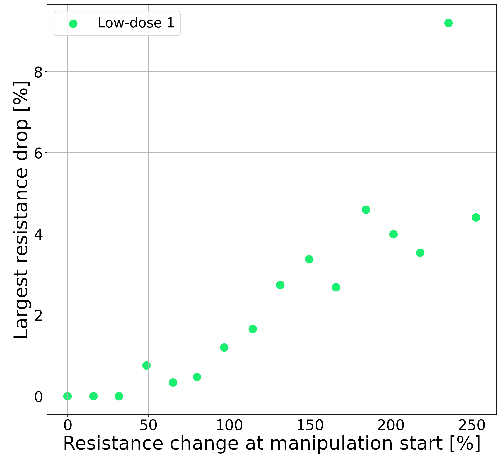}
    \caption{\scriptsize Initial resistance drop at each step during the stepped active manipulation in figure \ref{fig:stepped_active_manipulation}, as a function of $\rho_{\mathrm{Total}}$ obtained at the previous step. The drop magnitude increases with successive steps, after which the trend becomes irregular.}
    \label{fig:stepped_drop_depth}
\end{figure}

Furthermore, we note that the resistance drop increases with each successive step in the stepped active manipulation experiments. Figure \ref{fig:stepped_drop_depth} shows $\rho_{\mathrm{Drop}}$ as a function of $\rho(t)$ at the start of each manipulation step in figure \ref{fig:stepped_active_manipulation}. The initial drop deepens progressively with each step, before the behavior becomes chaotic. At present, we do not have a definitive explanation for this trend.

\subsection{Selecting a model for active resistance manipulation} \label{appendix:second_or_third_model}
We selected the polynomial model empirically by systematically comparing several candidate models for the resistance change $\rho(t) = R(t)/R(0) - 1$: a second-order polynomial, a third-order polynomial, an exponential, and a power law. Only the second- and third-order polynomials adequately captured the experimentally observed resistance evolution; their RMS errors are summarized in table \ref{tab:rmse_for_2nd_3rd}.
The mean difference shows that the performance of the third-order polynomial over the second-order polynomial is a few hundredths of percent better, at the cost of adding an additional degree of freedom into the model. We thus adopt the second-order polynomial as the simplest phenomenological model that represents the experimentally observed resistance increase over time, motivated by the fact that the model has fewer free variables at near-equal error.

\begin{table}[ht!]
    \centering
    \caption{ \scriptsize RMS error comparison between the second-order polynomial fit and a third-order polynomial fit for each resistance manipulation fit in figure \ref{fig:active_manipulation_room_temperature} in the same units as the figure, that is, \% resistance change. $V_a$ denotes the manipulation voltage. \label{tab:rmse_for_2nd_3rd}}
    \lineup
    \begin{tabular}{llllllllll}
    \br
    $V_a$ [\SI{}{\milli\volt}] & 750 & 800 & 850 & 900 & 925 & 950 & 1000 & 1050 & \\ 
    \br
    \textbf{Difference} & & & & & & & & & Mean \\
    \mr
    Low-dose 1        & 0.024 & 0.018 & 0.004 & 0.004  &        & 0.089 &        &        & 0.028 \\
    Low-dose 2        & 0.006 & 0.037 & 0.020 & 0.051  &        & 0.141 &        &        & 0.051 \\
    Medium-dose 1     &       &       &       & 0.0002 &        & 0.004 & 0.044   & 0.008 & 0.014 \\
    High-dose 1       &       & 0.003 & 0.006 & 0.003  & 0.004  & 0.005 & 0.00001 &       & 0.004 \\
    High-dose 2       &       & 0.003 & 0.003 & 0.0006 & 0.0003 & 0.004 & 0.0001  &       & 0.002 \\
    \br
    \textbf{2nd-order} & & & & & & & & & Mean \\
    \mr
    Low-dose 1        & 0.121 & 0.180 & 0.219 & 0.691 &       & 0.544 &       &       & 0.351 \\
    Low-dose 2        & 0.079 & 0.149 & 0.139 & 0.307 &       & 0.422 &       &       & 0.219 \\
    Medium-dose 1     &       &       &       & 0.078 &       & 0.092 & 0.123 & 0.129 & 0.105 \\
    High-dose 1       &       & 0.072 & 0.070 & 0.078 & 0.084 & 0.101 & 0.128 &       & 0.088 \\
    High-dose 2       &       & 0.066 & 0.067 & 0.088 & 0.083 & 0.086 & 0.095 &       & 0.081 \\
    \br
    \textbf{3rd-order} & & & & & & & & & Mean \\
    \mr
    Low-dose 1        & 0.097 & 0.162 & 0.214 & 0.686 &       & 0.455 &       &       & 0.323 \\
    Low-dose 2        & 0.072 & 0.112 & 0.119 & 0.256 &       & 0.281 &       &       & 0.170 \\
    Medium-dose 1     &       &       &       & 0.078 &       & 0.088 & 0.078 & 0.121 & 0.092 \\
    High-dose 1       &       & 0.069 & 0.064 & 0.075 & 0.080 & 0.097 & 0.128 &       & 0.085 \\
    High-dose 2       &       & 0.063 & 0.063 & 0.088 & 0.083 & 0.082 & 0.095 &       & 0.079 \\
    \end{tabular}
\end{table}

\subsection{Ohmic comparison of resistance change parameters}
Tables \ref{tab:alpha_and_beta_fit_low_dose} and \ref{tab:alpha_and_beta_fit_high_dose} present the fit results for $\alpha(V)$ and $\beta(V)$, multiplied by the respective $R(0)$ of each fitted trace, to convert the fractional units into \SI{}{\ohm\per\second} and \SI{}{\ohm\per\second\squared}, respectively. We denote these rescaled parameters as $\alpha'$ and $\beta'$. Similarly, table \ref{tab:exponential_fit_data} reports the resulting $\alpha_0'$ in units of \SI{}{\ohm\per\second}. For comparison, the characteristic voltage $V_0$ reported in table \ref{tab:exponential_fit_data_PERCENT} is also included.

\begin{table}[ht!]
    \centering
    \caption{ \scriptsize Extracted $\alpha'$ and $\beta'$ parameters for the \textit{low-dose} devices. \label{tab:alpha_and_beta_fit_low_dose}
    }
    \lineup
    \begin{tabular}{lllll}
    \br
            & \textbf{Low-}                & \textbf{Low-}                       & \textbf{Low-}                & \textbf{Low-}                       \\
            & \textbf{dose 1}                & \textbf{dose 1}                       & \textbf{dose 2}                & \textbf{dose 2}                       \\
    Voltage & $\alpha'$ [\SI{}{\milli\ohm\per\second}] & $\beta'$ [\SI{}{\micro\ohm\per\second\squared}] & $\alpha'$ [\SI{}{\milli\ohm\per\second}] & $\beta'$ [\SI{}{\micro\ohm\per\second\squared}] \\
    \mr
    \SI{750}{\milli\volt} & $2930 \pm 150$ & $-2600 \pm 500$   & $800 \pm 50$   & $-328 \pm 23$ \\
    \SI{800}{\milli\volt} & $6420 \pm 180$ & $-5700 \pm 800$   & $1560 \pm 40$   & $-1000 \pm 170$ \\
    \SI{850}{\milli\volt} & $12100 \pm 600$  & $-8230 \pm 250$ & $3920 \pm 17$ & $-2460 \pm 80$ \\
    \SI{900}{\milli\volt} & $22600 \pm 500$  & $-16500 \pm 2100$  & $8480 \pm 210$  & $-4660 \pm 80$ \\
    \SI{950}{\milli\volt} & $50300 \pm 400$  & $-33000 \pm 1500$  & $17960 \pm 120$ & $-7900 \pm 500$ \\
    \end{tabular}
\end{table}

{ \nobreak
\begin{table}[ht!]
    \centering
    \caption{ \scriptsize Extracted $\alpha'$ and $\beta'$ parameters for the \textit{medium-dose} and \textit{high-dose} devices. \label{tab:alpha_and_beta_fit_high_dose}
    }
    \lineup
    \begin{tabular}{lllllll}
    \br
            & \textbf{Medium-}                  & \textbf{Medium-}                         & \textbf{High-}                & \textbf{High-}                       & \textbf{High-}                & \textbf{High-}                       \\
            & \textbf{dose 1}                  & \textbf{dose 1}                         & \textbf{dose 1}                & \textbf{dose 1}                       & \textbf{dose 2}                & \textbf{dose 2}                       \\
    Voltage & $\alpha'$ [\SI{}{\milli\ohm\per\second}] & $\beta'$ [\SI{}{\micro\ohm\per\second\squared}] & $\alpha'$ [\SI{}{\milli\ohm\per\second}] & $\beta'$ [\SI{}{\micro\ohm\per\second\squared}] & $\alpha'$ [\SI{}{\milli\ohm\per\second}] & $\beta'$ [\SI{}{\micro\ohm\per\second\squared}] \\
    \mr
    \SI{800}{\milli\volt}  &                                     &                                   & $80.9 \pm 2.6$                        & $238 \pm 11$                      & $130 \pm 15$                         & $-174 \pm 7$                       \\
    \SI{850}{\milli\volt}  &                                     &                                   & $159 \pm 19$                         & $259 \pm 8$                        & $073 \pm 13$                         & $185 \pm 6$                        \\
    \SI{900}{\milli\volt}  & $1990 \pm 50$                       & $-2000 \pm 270$                       & $543 \pm 17$                         & $-76 \pm 8$                   & $255 \pm 11$                         & $-35 \pm 4$                       \\
    \SI{925}{\milli\volt}  &                                     &                                   & $602 \pm 15$                         & $153 \pm 6$                        & $446 \pm 5$                          & $26.6 \pm 1.3$                      \\
    \SI{950}{\milli\volt}  & $3460 \pm 30$                       & $-3090 \pm 120$                       & $780 \pm 230$                          & $-162 \pm 14$               & $632 \pm10$                         & $-173 \pm 5$                       \\
    \SI{1000}{\milli\volt} & $6762 \pm 23$                       & $-4710 \pm 100$                       & $2280 \pm 160$                          & $-950\pm60$                & $1700\pm100$                          & $-217\pm5$                       \\
    \SI{1050}{\milli\volt} & $10449\pm21$                        & $-4870\pm90$                        &                                     &                                   &                                     &                                   \\
    \end{tabular}
\end{table}

\begin{table}[ht!]
    \centering
    \caption{ \scriptsize Fit parameter $\alpha_0'$ and the corresponding characteristic voltage $V_0$. This voltage represents the additional voltage required for the linear rate of resistance change, to increase by a factor $e$.
    \label{tab:exponential_fit_data}}
    \lineup
    \begin{tabular}{llllll}
    \br
                                             & \textbf{Low-} & \textbf{Low-} & \textbf{Medium-}  & \textbf{High-}          & \textbf{High-}          \\
                                             & \textbf{dose 1} & \textbf{dose 2} & \textbf{dose 1}  & \textbf{dose 1}          & \textbf{dose 2}          \\
    \mr
    $\alpha_0'$ [\SI{}{\micro\ohm\per\second}] & $98.2 \pm 2.3$        & $21.5 \pm 1.3$        & $108.4 \pm 3.3$       & $0.0650 \pm 0.0005$                 & $0.0253 \pm 0.0005$                  \\
    $V_0$ [\SI{}{\milli\volt}]               & $72.5 \pm 2.4$        & $70.4 \pm 3.4$        & $91.4 \pm 4.3$        & $57.5 \pm 5.0$                  & $55.3 \pm 3.1$                  \\
    \end{tabular}
\end{table}
}

\section{Residual analysis between relaxation models} \label{appendix:log_vs_power_in_relaxation}
To characterize the relaxation growth, we compare the logarithmic model in equation (\ref{eq:log_fit}) with a power-law description,
\begin{equation}
    y(t) = a + c \cdot t^{d} .
\end{equation}
We evaluate the mean and variance of the residuals $\Delta k,~ \Delta m$ between the fitted curves and the measured data. For the logarithmic model, the mean residuals are $\langle \Delta k \rangle = -1.42 \cdot 10^{-5}$ and $\langle \Delta_m \rangle = -8.59 \cdot 10^{-6}$, with variances $2.04 \cdot 10^{-5}$ and $1.63 \cdot 10^{-3}$, respectively. In contrast, the power-law model yields larger residuals, with mean values $\langle \Delta k \rangle = -9.16 \cdot 10^{-5}$ and $\langle \Delta m \rangle = -1.06 \cdot 10^{-4}$ and variances $5.16 \cdot 10^{-5}$ and $2.18 \cdot 10^{-3}$. The consistently smaller residuals and variances indicate that the relaxation growth is better described by a logarithmic dependence than by a power law. In all fits, we enforced a minimum bound for $a$ being $1$ corresponding to $\rho_\mathrm{Total}(t_\mathrm{Stop}) / \rho_\mathrm{Active}(t_\mathrm{Stop}) = 1$, as no relaxation has yet occurred at the stopping time.

\section{A thermally activated process} \label{section:thermal_process}
The term \textit{electrical annealing} is commonly used to describe the resistance manipulation process studied in this work. However, the terminology \textit{annealing} can be misleading, as there is strong evidence that the underlying physical process does not primarily rely on heating the junction to conventional annealing temperatures. Instead, the experimental picture that emerges is that electrical tuning is fundamentally a thermally activated process, in which the electric field plays an essential role.

To demonstrate how the manipulation process is influenced by temperature, we performed an ad-hoc experiment. We performed resistance manipulation at room temperature, after which the junction was allowed to relax while submerged in liquid nitrogen. The experiments were carried out in an aluminum enclosure, surrounded by a styrofoam insulating layer, mounted onto the manual probe station. The enclosure was designed to be filled with liquid nitrogen through a dedicated inlet. A test chip containing \textit{low-dose} Josephson junctions was glued to the bottom of the box using {\nobreak BF-6} Bakelite phenol-formaldehyde adhesive, and a {\nobreak Si-540} silicon diode thermometer was mounted nearby for temperature monitoring via a spare channel of the source-measurement unit. Electrical connections to the sample were made by inserting the manual station's probe needles through a large opening in the box lid, which also allowed nitrogen boil-off to escape and thereby suppressed condensation on the sample. A secondary gas inlet enabled controlled warmup using room-temperature nitrogen gas and provided a slight overpressure to mitigate moisture accumulation. Significant thermal contraction and expansion of the probe needles during cooldown and warmup required repeated manual realignment to maintain electrical contact with the sample.

\begin{figure}[ht!]
    \centering
    \includegraphics[width=1.00\textwidth]{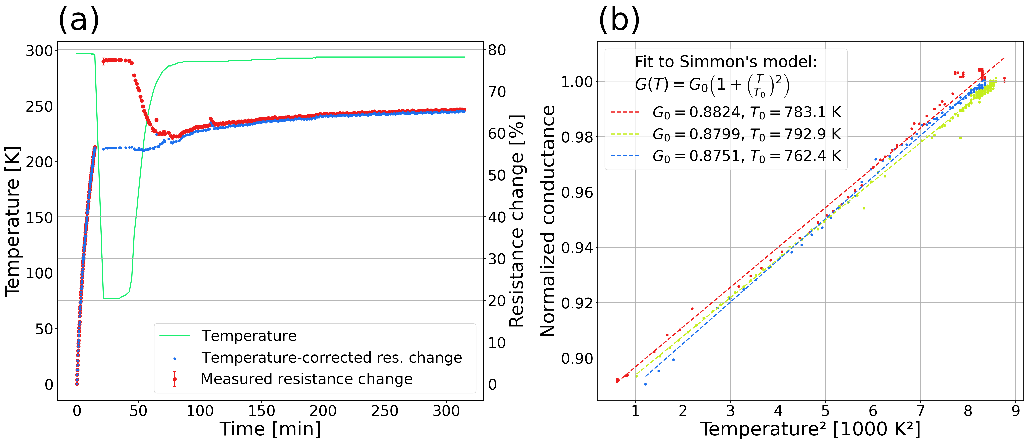}
    \caption{ \scriptsize (a) Electrical tuning of a Josephson junction with the relaxation performed in liquid nitrogen.
    No resistance change is observed while the junction remains cold, a result further emphasized by comparison with the temperature-corrected resistance trace shown as blue points. (b) Determination of the average characteristic temperature $\langle T_0 \rangle$ and normalized conductance $\langle G_0 \rangle$ used for the Simmons-model correction in equation (\ref{eq:room_temperature_equivalent}), extracted from three liquid-nitrogen cooldowns in which temperature and conductance were monitored for three \textit{low-dose} junctions}
    \label{fig:ln2_experiment}
\end{figure}

Figure \ref{fig:ln2_experiment}(a) shows an electrical tuning experiment of a $183$-day-old \textit{low-dose} junction, where the device is submerged in liquid nitrogen immediately after resistance manipulation, so that relaxation occurs at lower temperatures. Remarkably, no resistance change is observed while the junction remains cold.

To further emphasize that the relaxation-induced resistance change halts at low temperature, we normalize the measured resistance to its room-temperature equivalent using an approximation to Simmon's model for tunneling through a thin insulator barrier,
\begin{equation} \label{eq:simmons_model}
    G(T) = G_0 \cdot \left(1 + \left( \frac{T}{T_0} \right)^2 \right) ~~,
\end{equation}
\noindent where $G_0$ and $T_0$ are the characteristic conductance and characteristic temperature, respectively. This model is expected to be applicable since the junction never becomes superconducting during the experiment, and behaves as a metal-insulator-metal junction with a negative temperature coefficient. We find a conversion between the room-temperature conductance and the cryogenic conductance, 
\begin{equation} \label{eq:room_temperature_equivalent}
    G_{\text{RT}}(T) = G_{\text{Cryo}}(T) \cdot \left( \frac{G_0 \cdot \left(1 + \left( \frac{\SI{297}{[\kelvin]}}{T_0} \right)^2 \right)}{G_0 \cdot \left(1 + \left( \frac{T}{T_0} \right)^2 \right)}\right) ~~,
\end{equation}
\noindent where $1/G_\mathrm{Cryo} = R$ is the resistance of the junction as we measure it in-situ.
Both $G_0$ and $T_0$ are determined experimentally in figure \ref{fig:ln2_experiment}(b), where three \textit{low-dose} junctions are cooled down, while their resistance and temperature are continuously monitored without performing any electrical tuning. Fitting these three datasets to equation (\ref{eq:simmons_model}) yields an averaged normalized conductance \mbox{$\langle G_0 \rangle = 0.8791~$}, and an averaged \mbox{$\langle T_0 \rangle =~$\SI{779.5}{\kelvin}}. $G_0$ would, in principle, correspond to the zero-temperature conductance of the junction; its inverse gives
\begin{equation}
    1/\langle G_0 \rangle =R/R_N = 1.1375  ~~,
\end{equation}
\noindent in close agreement with the $R/R_N$ factor $1.1385$ used throughout this work.
Using $\langle T_0 \rangle$ in equation (\ref{eq:room_temperature_equivalent}), figure \ref{fig:ln2_experiment}(a) shows the resulting temperature-corrected resistance, confirming that resistance relaxation is effectively frozen out at cryogenic temperature. Moreover, resistance relaxation resumes as the device thaws. It should also be mentioned that electrical tuning itself is known to speed up at elevated temperatures \cite{Pappas2024}

We initially suspected that the resistance change observed during active manipulation arises from localized Joule heating at the junction's metal-insulator interface, caused by power dissipation $P = V_{\mtext{drop}} \cdot I$. In \cite[figure 4.9]{Toselli2024_thesis}, this hypothesis was tested by simulating the localized heating in a cross-style Josephson junction \cite{Potts2001} under DC manipulation. For a representative case of $V_a = \SI{0.9}{\volt}$ applied for \SI{10}{\minute}, the junction dissipates approximately \SI{110}{\micro\watt} at the Al-AlO$_\mtext{x}$ interface. Assuming that $10$\% of the junction area participates in tunneling \cite{Zeng2015}, the resulting tunnel current is \SI{120}{\micro\ampere}, corresponding to a simulated local temperature rise of $\sim\! \SI{4}{\kelvin}$. This result is an order of magnitude below the levels required for conventional thermal annealing, and is also consistent with a recent study reporting $\sim\! \SI{8}{\kelvin}$ local heating in comparable junctions \cite{Kennedy2025}.

These results raise the question of whether the relaxation process is fundamentally distinct from, or closely related to, natural aging.
Aging studies of Josephson junctions \cite{Koppinen2007, Bladh2005} show that aging slows at lower temperatures, with their glassy relaxations showing the same trend \cite{Nesbitt2007}.
Both aging and relaxation also share in common that subjecting the sample to sufficient heat halts further aging \cite{Koppinen2007}, as well as the ability to change resistance from electrical tuning \cite{alegria2025abaa}.
While there may be other interpretations that we cannot rule out, our measurements in this work are consistent with the following model. The tunnel barrier can be described by an energy landscape containing many metastable local minima. During natural aging, thermal fluctuations at room temperature provide a finite probability for atoms, defects etc. to overcome local activation barriers and relax toward lower-energy configurations, thereby increasing the junction resistance. Electrical manipulation could then temporarily modify the energy landscape, the electric field then lowers certain activation barriers sufficiently for room-temperature thermal energy to drive transitions that would otherwise occur only on very long timescales through natural aging.
This model is also consistent with the observations reported in \cite{alegria2025abaa} where the claim was that subjecting the junctions to thermal treatment, lowers the ability to perform ABAA, and even removing any ability to perform ABAA if heating junctions to sufficiently high temperature. Such behavior would be expected if the system approaches a global energy minimum by thermal fluctuations, leaving fewer accessible metastable states available for electrically induced rearrangement. We perform electrical tuning on samples that have aged for multiple months, further suggesting that the active manipulation accesses configurations that are not reached through natural aging alone. In active manipulation, the extracted characteristic voltages $V_0$, in equation (3), have values in the range \SI{55}{}$-$\SI{91}{\milli\volt}; these values are notably comparable to, but larger than, the thermal voltage at room temperature $V_T \approx \SI{25}{\milli\volt}$. This observation is consistent with the interpretation outlined above. The observation reported in \cite{Pappas2024} that alternating the pulse polarity enhances the range and rate of resistance change could be explained due to enabling transitions out of metastable configurations that become energetically unfavorable under the reversed field direction.

Recent hypotheses \cite{Pappas2024, Wang2024} speculate that electrical tuning is related to field-assisted oxidation through the Cabrera-Mott mechanism \cite{Cabrera1949}. Within this frame, electrical tuning accelerates the intrinsic relaxation dynamics of the junction not through heating, but by providing an electric-field-driven impulse that promotes ionic rearrangement toward a more stable configuration. However, this interpretation must be reconciled with experimental results showing that Cabrera-Mott oxidation increases as the temperature is lowered \cite{Cai2012}, in contrast to observations that electrical tuning speeds up with increased temperature.

\section{Dielectric breakdown} \label{section:dielectric_breakdown_results}
\begin{figure}[ht!]
    \centering
    \includegraphics[width=1.00\textwidth]{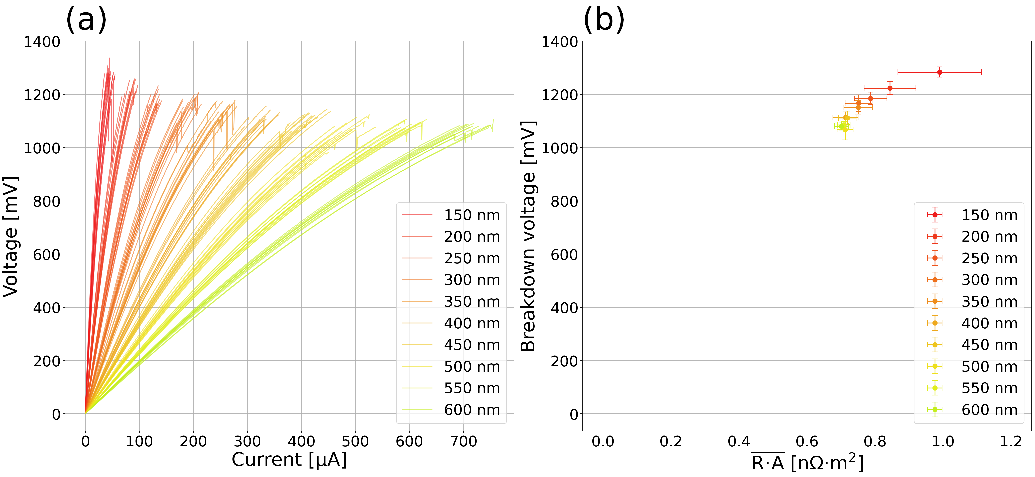}
    \caption{ \scriptsize Current-ramped dielectric breakdown measurements performed on sacrificial Josephson junctions from the \textit{high-dose} wafer. (a) Individual I-V curves for $119$ junctions of different sizes. (b) Breakdown voltage as a function of the mean resistance-area product. The legends show the electrode width of the junctions.}
    \label{fig:breakdown_image}
\end{figure}

Figure \ref{fig:breakdown_image} shows dielectric breakdown characteristics of \textit{high-dose} junctions. In figure \ref{fig:breakdown_image}(a) the automated probe station was used to apply a current ramp to $119$ junctions with areas ranging from $\SI{150}{\nano\meter}~\text{x}~\SI{150}{\nano\meter}$ to $\SI{600}{\nano\meter}~\text{x}~\SI{600}{\nano\meter}$, while monitoring the voltage across them. The breakdown voltage $V_{\mathrm{break}}$ is defined as the highest voltage measured before the resistance dropped below \SI{260}{\ohm}, a threshold determined empirically as the upper bound for the short-circuit resistance of a failed junction. For junctions with electrode widths between \SI{400}{\nano\meter} and \SI{600}{\nano\meter}, we consistently observed $V_{\mathrm{break}} \approx \SI{1.1}{\volt}$. As shown in figure \ref{fig:breakdown_image}(b), smaller junctions exhibit increased scatter in both breakdown voltage and resistance-area product, reflecting increased measurement uncertainty and device-to-device variability at reduced dimensions. These results indicate that, for each wafer, dielectric breakdown measurements on sacrificial junctions are essential for establishing safe upper bounds on the voltage amplitude used during electrical tuning.

\subsection{Survival duration during tuning}
\begin{figure}[ht!]
    \centering
    \includegraphics[width=0.50\textwidth]{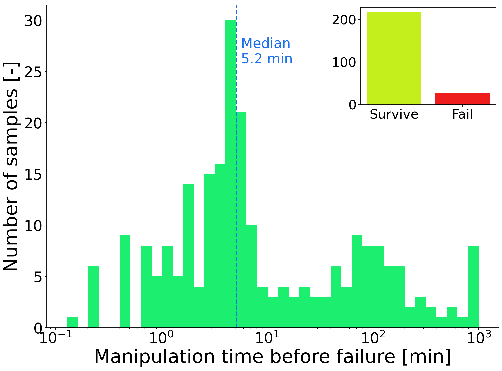}
    \caption{ \scriptsize Survival duration analysis of all junction variants that failed at any point during active manipulation with $V_a = \SI{0.85}{\volt}$. For a junction failing in this work, the median survival time is \SI{5.2}{\minute}. Across all junction variants studied, the overall survival rate is $89.3\%$, with the 10-percentile being \SI{1.0}{\minute} and the 90-percentile being \SI{167.0}{\minute}.}
    \label{fig:survival}
\end{figure}

Figure \ref{fig:survival} shows a histogram of the survival duration for $n=243$ junctions that at any point broke during manipulation, using the regular manipulation procedure outlined in section \ref{section:active_resistance_manipulation}. All failures resulted from a short circuit through the junction, defined as measuring a junction resistance of less than \SI{260}{\ohm}. The data includes results from stepped active manipulation experiments, where we counted only the first step of the manipulation since this step is identical to a regular manipulation. We achieve a survival rate of $89.3\%$.  We believe that the electrical tuning protocol requires identifying a safe tuning voltage for junctions of that particular wafer. In our work, selecting $V_a = \SI{0.85}{\volt}$ for all junction variants, while consistent across all variants studied, is not a one-size-fits-all; as was shown in table \ref{tab:exponential_fit_data_PERCENT}, the rate of resistance change spans multiple orders of magnitude even for the same manipulation voltage.

\clearpage

\bibliographystyle{iopart-num.bst}
\bibliography{master.bib}

\end{document}